\documentclass[12pt,preprint]{aastex}

\begin{document}
\def\cc{}
\def\c2{}

\title{Cosmological Feedback from High-Redshift Dwarf Galaxies}

\author{Akimi Fujita$^{1,2,3}$, Mordecai-Mark Mac Low$^{1,3}$, Andrea
Ferrara$^4$ and Avery Meiksin$^5$}  

\affil{$^1$Astronomy Department, Columbia University, 550 West 120th
Street, New York, NY 10027, USA} 
\affil{$^2$Physics Department, University of California, Santa Barbara, 
CA 93109, USA}
\email{fujita@physics.ucsb.edu}
\affil{$^3$Astrophysics Department, American Museum of Natural
History, Central Park West at 79th Street, New York 10024-5192, USA}
\email{mordecai@amnh.org} 
\affil{$^4$SISSA/International School for Advanced Studies, 
Via Beirut 4, 34014 Trieste,  Italy}
\email{ferrara@arcetri.astro.it} 
\affil{$^5$Institute for Astronomy,
University of Edinburgh, Blackford Hill, Edinburgh EH9 3HJ}
\email{aam@roe.ac.uk}

\begin{abstract}
We model how repeated supernova explosions in high-redshift dwarf
starburst galaxies drive superbubbles and winds out of the
galaxies. We compute the efficiencies of metal and mass ejection and 
energy transport from the galactic potentials, including the effect of 
cosmological infall of external gas. The starburst bubbles quickly blow 
out of small, high-redshift, galactic disks, but must compete with the 
ram pressure of the infalling gas to escape into intergalactic space.
We show that the assumed efficiency of the star formation rate dominates the
bubble evolution and the metal, mass, and energy feedback efficiencies. 
With star formation efficiency $f_{*}=0.01$, the ram pressure of
infall can confine the bubbles around high-redshift dwarf galaxies
with circular velocities $v_c\ga52$ km s$^{-1}$. 
We can expect high metal and mass ejection efficiencies, and moderate 
energy transport efficiencies in halos with $v_c\approx30-50$ km s$^{-1}$ 
and $f_*\approx0.01$ as well as in halos with $v_c\approx100$ km s$^{-1}$ 
and $f_*\gg 0.01$. Such haloes collapse successively from 1--2$\sigma$ peaks 
in $\Lambda$CDM Gaussian density perturbations as time progresses.
These dwarf galaxies can probably enrich low and high-density regions 
of intergalactic space with metals to $10^{-3}$--$10^{-2}$ Z$_{\odot}$
as they collapse at $z\approx8$ and $z\la5$ respectively.  
They also may be able to provide adequate turbulent energy to prevent
the collapse of other nearby halos, as well as to significantly broaden 
Lyman-$\alpha$ absorption lines to $v_{rms}\approx20$--40 km s$^{-1}$. 
We compute the timescales for the next starbursts if gas freely falls back 
after a starburst, and find that, for star formation efficiencies as low as 
$f_{*} \la 0.01$, the next starburst should occur in less than half the Hubble 
time at the collapse redshift. This suggests that episodic star formation may 
be ubiquitous in dwarf galaxies.
\end{abstract}

\section{Introduction}
\label{sec:intro}

\subsection{Overcooling and Angular Momentum Problems}
The theory of hierarchical structure formation in a universe dominated
by cold dark matter (CDM) predicts that galaxies assembled
from mergers of dark matter halos that gravitationally collapsed from
primordial fluctuations.  Within the same framework, the first
luminous structures to form have sub-galactic masses, given by the
Jeans mass of the gas.  This in turn is determined by the cooling
mechanisms dominant during each cosmological era, since CDM models do
not define a lower limit to the scale of inhomogeneities in the
overall mass distribution.  Protogalactic gas clouds of mass
$M_h\sim10^{9}$~M$_{\odot}$
begin to collapse as $\sim3\sigma$ peaks
around redshift $z\sim10$ and become dwarf galaxies.  

The hierarchical scenario
predicts the formation of numerous dwarf-sized halos at
high redshift.  In these halos, the dissipative collapse of gas is
very efficient, since the cooling time
$\tau\propto\rho^{-1}\propto(1+z)^{-3}$, where $\rho$ is the mass
density.  This is often referred to as the ``overcooling problem''
(White \& Rees 1978; White \& Frenk 1991).  Overcooling leads to a
prediction of far more dwarf galaxies than observed, as well as to an
angular momentum problem.  When dense, cooled, protogalactic gas
clouds or dwarf galaxies interact within a larger system such as the
halo of a present-day disk galaxy, dynamical friction transfers their
orbital angular momentum to the surrounding dark matter halo.  The
result is a galaxy with too much mass, rotating too slowly, with
angular momentum deficient by a factor of $\sim25$ relative to that
observed (e.g.\ Katz \& Gunn 1991; Navarro \& Benz 1991; Navarro \&
White 1994).

Both of these problems can be resolved if a mechanism can be found to
keep the gas diffuse until the peak of present-day disk ($L_*$) galaxy
formation.  Supernova feedback from young stars is one commonly
invoked mechanism to do this.  A number of cosmological hydrodynamic
simulations have demonstrated that stellar feedback plays a major role
in the formation of galaxies (Katz 1992; Navarro \& White 1993; Yepes
et al.\ 1997; Gerritsen \& Icke 1997; Hultman \& Pharasyn 1999;
Scannapieco et al.\ 2001; Thacker \&
Couchman 2001; Sommer-Larsen et al.\ 2002; Springel \& Hernquist
2003).  However, the qualitative outcomes of simulations vary
drastically with the feedback scheme applied.  Recently, smoothed
particle hydrodynamic (SPH) simulations by Scannapieco et al.\ (2001)
and Springel \& Hernquist (2003) show that overcooling in low-mass halos
is significantly suppressed by the galactic outflows that they
model, since the outflows slow down the accretion of gas in the halos
and also strip baryons from neighboring halos ("baryonic stripping").
The latest high-resolution ($\sim10^{7}h^{-1}$M$_{\odot}$ per gas
particle) SPH simulations by Thacker and Couchman (2001) and
Sommer-Larsen et al.\ (2002) show that their treatment of thermal
feedback with suppressed cooling succeeds in reproducing a present-day
disk galaxy with angular momentum that is deficient only by a factor
of a few.

However, these cosmological simulations still fail by orders of
magnitude to resolve the physics of star formation and feedback.  The
coupling of feedback energy with interstellar gas through shocks, the
formation of galactic outflows, and the coupling of the outflow energy
with halo gas are not physically represented, 
especially in SPH simulations with resolution at the galactic scale
far too low to accurately resolve shocks.  Therefore, it is important
to model the collective action of multiple supernovae and study the
formation of superbubbles and galactic winds in single dwarf galaxies,
using high-resolution, hydrodynamic simulations, in order to compute
the feedback efficiencies (e.g.\ Mac Low \& Ferrara 1999; hereafter
MF99). Such models must also include cosmological infall of gas at
high redshift, because the ram pressure of the infall may influence
the evolution of the superbubbles and the galactic winds.
  
%

\subsection{Intergalactic Metals}
Observations of Ly$\alpha$-absorbing clouds reveal the presence of metals.
These clouds are thought to be regions of enhanced intergalactic
medium (IGM) density distant from protogalaxies (see review by Rauch
1998).  To transport metals into these regions, some mechanism such as
supernova-driven galactic outflows must act.  The observations
show that Lyman forest clouds with neutral hydrogen column density with
$\log N(HI)>14$
are metal enriched to $Z\approx 10^{-3}-10^{-2}$ Z$_{\odot}$ 
(Cowie et al.\ 1995; Songaila \& Cowie 1996; Ellison
et al.\ 2000), and that the metallicity remains roughly constant
throughout the redshift range $1.5 < z <5.5$ (Songaila 2001).  It
is also interesting to note that $\log N(HI)\approx 14$ at $z\approx3$
marks the transition between continuous filamentary structures and
voids (e.g.\ Zhang et al.\ 1998).

The main candidates for the polluters of the IGM are starburst
dwarf galaxies.  The absence of turbulent motions observed in
low-density Lyman $\alpha$ clouds at $z\approx3$ (Rauch et al.\ 2001b)
suggests that metal enrichment was completed at very early times
($z \gg 5$), when the physical volume of the universe was
smaller, and contained the numerous dwarf galaxies predicted by the
CDM model of galaxy formation (Rauch et al.\ 2001b; Madau, Ferrara, \&
Rees 2001; Scannapieco, Ferrara, \& Madau 2002).
{\c2 On the other hand, the median Doppler parameters are significantly larger
in 
the 
Lyman $\alpha$ forest than predicted by cosmological simulations, suggesting 
additional energy injection provided by late He{\sc II} reionization 
or supernova-driven winds (Meiksin, Bryan, \& Machacek 2001).  
In C{\sc iv} systems with higher column densities, a substantial
velocity scatter over length scales of a few hundred parsecs is observed, 
however,
suggesting} they have been influenced by galactic feedback more recently
(Rauch et al.\ 2001a; see Rauch 2002 for summary).  Galactic outflows
both at very high redshift and at lower redshift $z\sim3$ seem to play
significant roles in enriching various regions of the IGM.

{\cc Cosmological simulations have suggested that the IGM at $z\approx3$
can be metal-enriched to $Z=10^{-3}$--$10^{-2}$~Z$_{\odot}$
by merging of 
or by outflows from dwarf galaxies (Gnedin 1998; Theuns et al.\ 2002;
Thacker et al.\ 2002) without 
dynamically disturbing the observed, low-density, Ly$\alpha$ clouds.}
  These simulations have the same
problem as earlier that the feedback schemes differ based on different
assumptions about the coupling of supernova energy to the
surroundings, and the simulations fail badly to resolve star formation
and the dynamics of bubbles.

\subsection{Starburst Wind Models}

The effects of repeated supernova explosions from starbursts appear to be a
central piece of physics necessary for understanding the role of
stellar feedback in both galaxy formation and IGM pollution.
The effects of starbursts from dwarf galaxies on the surrounding
interstellar medium have been studied in the past both analytically
and numerically (Mathews \& Baker 1971; Larson 1974; Saito 1979; Dekel
\& Silk 1986; De Young \& Heckman 1994; Silich \& Tenorio-Tagle 1998;
MF99; D'Ercole \& Brighenti 1999).  For example, MF99 studied the
effects of repeated supernovae on local dwarf galaxies with gas mass
$M_{g}=10^{6}$--$10^{9}$~M$_{\odot}$ using hydrodynamic simulations,
and made a parameter study of mass and metal ejection efficiencies.
The main results of MF99 are that mass loss is very inefficient except
in the lowest mass halos, while a substantial fraction of the hot,
metal-enriched gas escapes from any of the potentials they studied.
These results were confirmed by D'Ercole \& Brighenti (1999) with a
better treatment of thermal conduction.  Recently, Mori, Ferrara, \&
Madau (2002) studied the same problem, but in a spherical,
non-rotating, galaxy embedded in a halo with $M_h=10^{8}h^{-1}$
M$_{\odot}$ at $z=9$.  
They find that, with
star formation efficiency $f_* =0.1$, about 30\% of the available
supernova energy is transferred to the surrounding gas as kinetic
energy. The starburst bubbles expand over $\gg 10R_v$, where $R_v$ is
the virial radius of the halo, comparable to the mean proper distance
between neighboring low-mass systems. Their results support the
suggestion that IGM metal enrichment occurs early. 
{\cc The recent study by Wada \& Venkatesan (2003) improves 
on Mori et al.\ (2002)
by precalculating a thin, dense disk with an inhomogeneous ISM 
in a given halo potential and by including self-gravity. 
They perform two three-dimensional models of a $10^8$~M$_{\odot}$
galaxy with star formation efficiency $f_*=0.14$ and~0.014 in our
terms.}


We follow the work of MF99 and extend their study to high
redshift. 
{\cc Although the approachs taken by Mori, Ferrara, \& Madau
(2002) and Wada \& Venkatesan (2003) are more realistic in some ways, 
the former neglects rotational flattening of the galaxy, 
and both are far more expensive computationally.}
The MF99 approach enables us to do a parameter study
on the feedback process by exploring a wide range of galaxy masses,
formation redshifts, and star formation efficiencies. 
{\cc However, note that our study is limited to axisymmetric 
geometry
and also to single star formation sites at the centers of disks with
a smooth ISM.}

We perform careful, high-resolution models of the result of multiple supernova
explosions in single dwarf galaxies at high redshift, using a
hydrodynamic code, ZEUS-3D (Stone \& Norman 1992; Clarke 1994), and
compute the efficiency of metal and mass ejections and energy transport from the
galactic potentials.  We include the evolving dark matter halo
potentials, and the cosmological infall of halo gas
computed with a one-dimensional hydrodynamic code (Meiksin 1994).  The
bubbles in MF99's local galaxies were solely governed by their
interaction with the interstellar medium (ISM), and so high metal
ejection efficiencies were guaranteed once the bubbles blew out of the
galactic disks.  However, the bubbles in our high-redshift galaxies
must still fight the ram pressure of the infalling halo gas after
blowout before the metals, swept-up mass, and energy escape to
intergalactic space. We also compute the accretion timescales for any
mass swept up by the bubbles, but bound by the potential to fall back
to the center.

We model systems that we think likely to host a starburst. 
It is not certain whether first
generation systems 
were efficient in ejecting metals and transporting energy to the IGM,
because an early cosmic UV background suppreses the formation of stars inside
the systems by photodissociating their only cooling agent, molecular hydrogen 
(Haiman, Abel, \& Rees 2000), 
{\cc or by raising the entropy floor (Oh \& Haiman 2003)}
However, later systems that collapsed due to atomic hydrogen cooling may have been
more robust allowing starbursts to occur (Madau, Ferrara, \& Rees
2001; Scannapieco, Ferrara, \& Madau 2002; Oh \& Haiman 2002).

We choose seven of these second generation systems, with
$5\times10^{8}$~M$_{\odot} \le M_h \le 5\times10^{10}$ M$_{\odot}$ at $3\le
z\le 13$ with star formation efficiencies $f_{*}=0.001$, 0.01,
and 0.1.  Our goal is to develop a generally useful description of
supernova feedback in terms of metal ejection efficiencies
$\xi_{metal}$, mass ejection efficiencies $\xi$, and energy transport
efficiencies $\zeta$, that can be widely used in large-scale
cosmological simulations for the study of galaxy formation and metal
enrichment.  However, it will be shown in this paper that
parameterization of feedback is very difficult, and that we eventually
need a higher-resolution, adaptive mesh refinement, cosmological
simulation with a realistic treatment of star formation to do so as
reliably as we might wish. Nonetheless, we are able to reach several key conclusions
with the present set of computations. 

The paper is organized as follows.  In \S\ref{sec:predict} we
analytically predict the effects of ram pressure on the confinement of
the starburst bubbles. In \S\ref{sec:numer} we discuss our numerical
methods, including our models of cosmological infall
(\S\ref{sub:halo}), starburst blowout (\S\ref{sub:2D}), and the
tracer field we use to follow metal-enriched gas in the blowout model
(\S\ref{sub:tracer}).  In \S\ref{sec:disk} we describe our choices for
galaxy size (\S\ref{sub:galaxies}), disk model (\S\ref{sub:disks}), and
star formation feedback (\S\ref{sub:starbursts}).
We give our results in \S\ref{sec:results}: metal and mass ejection, and energy
transport efficiencies in \S\ref{sub:feedback}, the effects of external
pressure on the bubble evolution in \S\ref{sub:press}, and the
timescales for material to fall back in \S\ref{sub:time}.  Our
conclusions follow in \S\ref{sec:concl}.  Throughout the paper, we
apply $\Lambda$CDM cosmology with $\Omega_{0}=0.37$,
$\Omega_{\Lambda}=0.63$, $\Omega_{b}=0.05$, $h=0.7$, $\sigma_8=0.8$ 
and a primordial composition of H:He=12:1.


\section{Pressure Confinement}
\label{sec:predict}
We first try to estimate analytically how the ram pressure of
infalling gas affects the evolution of a starburst bubble, in order to
estimate the ranges of halo mass $M_{h}$, redshift $z$, and star
formation efficiency $f_{*}$ over which the ram pressure confinement
of a bubble suppresses feedback.

In this section, we consider a bubble in a power-law halo with gas density $\rho=\rho_0
r^{-\alpha}$. Isothermal halos with constant angular velocity have
$\alpha =2$. We take the bubble to be driven by a mechanical
luminosity $L_m$ for the 50~Myr lifetime of the least massive star
able to go supernova after an instantaneous starburst, and then to
evolve freely thereafter.  A bubble big enough to interact with the
halo will blow out of the highly stratified, exponential atmosphere of
a high-redshift galactic disk very quickly.  After it blows out,
its evolution is largely determined by the 
infalling halo gas. Since the bubble accelerates as it blows out of the
exponential ISM (Kompaneets 1960), it is difficult to exactly predict
its expansion. Therefore, we make the simplifying assumption
that the bubble has reached the power-law
halo by 50~Myr, which is supported by our detailed numerical calculations
described in section~\ref{sec:results} below. Thereafter, we compare the interior pressure of the bubble
with the ram pressure of the infalling halo gas.
 
After the starburst ends at a time $t_f=50$~Myr, the bubble continues
to expand due to the residual energy of supernovae stored in the hot
interior gas and in the shell.  However, we do not know how much of
the injected energy $L_m t_f$ remains in the interior, rather than
being dissipated by radiative cooling enhanced by shell fragments mixing
with the interior gas, and how much of it is transferred to the kinetic
and thermal energy of the swept-up shell. Therefore, we assume a fraction
$\nu$ of the injected energy $L_m t_f$ drives the bubble after
$t_f$.

From the Rankine-Hugoniot jump conditions, we have
\begin{equation}
P=\frac{2}{\gamma+1} \rho u^2_{1},
\label{pressure}
\end{equation}
where $P$ is the pressure behind the shock front,
$\rho=\rho_{0}R^{-\alpha}_s$, the radius of the shock is $R_{s}$, and
$u_{1}$ is the velocity ahead of the shock, in a frame of reference
traveling with the shock velocity $\dot{R_{s}}$.  The preshock
velocity 
%
is
%
\begin{equation} u_{1}=\dot{R_s} + v_{in}(R_s),
\end{equation}
where the infall velocity of gas is $ v_{in}(R_s)$.  We assume that
the pressure $P$ behind the shock equals the mean pressure $\bar{P}$
of the gas within the spherical volume enclosed by $R_{s}$, and a
fraction $\nu$ of the total injected energy $L_m t_f$ gives the
pressure (Spitzer 1978).  
{\cc (It is equivalent to setting $\zeta=1$ 
in the Kompaneets approximation; however, in reality, 
it is not always true that $P =\bar{P}$: 
for example, $P \simeq 2\bar{P}$ in a Sedov blast wave
solution [Sedov 1959])}.  We can then derive the shock velocity,
\begin{equation}
\dot{R}_{s} = 0.46\left(\frac{\nu L_m
t_f}{\rho_{0}}\right)^{1/2} R^{(\alpha-3)/2}_s - v_{in}(R_{s}),
\label{rsdt}
\end{equation} 
where we have taken $\gamma=5/3$ for an adiabatic bubble shock (but
the coefficient varies by only 15\% for a radiative shock with $\gamma
= 1$), and we assume the thermal pressure of the IGM
to be much smaller than its ram pressure.  The shock velocity driven
by the mechanical luminosity $L_m$ prior to $t_f$ exceeds that given
by equation~(\ref{rsdt}), but the bubble quickly decelerates without
pressure support from the central energy source.
 
We approximate the infall velocity of gas,
\begin{equation}
v_{in}(r)=(2GM_{h}/r)^{1/2},
\label{vin}
\end{equation}
as that of a free fall onto a point mass with the mass $M_h$ of the
halo, as long as the radius $r$ exceeds the virial radius $R_v$ (given
below by Eq.~[\ref{rv}]).  We completely neglect the angular momentum
of the infalling gas, so $v_{in}$ is an upper limit. 
{\cc The bubble will stall when $\dot{R_s}=0$ by ram pressure exerted by 
the infalling gas.}
Then, we can derive
\begin{equation} 
v_{in}(R_s)\ge 0.46 \left(\frac{\nu L_m t_f}
{\rho_{0}}\right)^{1/2} R^{(\alpha-3)/2}_s,
\label{rp}
\end{equation}
as the criterion for ram-pressure confinement.  Both sides of the
equation are functions of the bubble radius $R_{s}$, but we can
eliminate the radial dependence in an isothermal halo with $\alpha=2$
and compare both sides of the equation directly without knowing where
the bubble is.  Note that the left side of the equation is just under
the escape velocity of a halo $v_{esc}\approx(GM_h/R_v)^{1/2}$ at
$r\ga R_v$, and {\cc the right side is the shock velocity of a
bubble if there is no infalling gas.}

The coefficient $\rho_0$ for the 
gas density of a spherically symmetric, power-law medium is 
\begin{equation}
\rho_{0}=M_{h}\left(\frac{\Omega_{b}}{\Omega_{0}}\right) 
\left(\frac{3-\alpha}{4\pi}\right) R_v^{\alpha-3}, 
\label{rho0}
\end{equation}
when a given total halo mass $M_{h}$ is found within the virial radius $R_v$. 
The mechanical luminosity $L_m$ can be expressed 
in terms of star formation efficiency $f_{*}$ as
%
%
\begin{equation}
L_m=(3.6\times10^{31} \mbox{ erg s}^{-1}) \left(\frac{M_{h}}{\mbox{ M}_{\odot}}\right)
\left(\frac{f_{*}}{0.01}\right),
\label{lmech}
\end{equation}
%
based on the Starburst 99 model (Leitherer et al.\ 1999) described in \S\ref{sec:disk}.

Substituting equations~(\ref{vin}), (\ref{rho0}), and (\ref{lmech})
into equation~(\ref{rp}) with $\alpha=2$, we find that
external ram pressure can confine a bubble at a given redshift $z$
when the total halo mass
\begin{equation}
M_{h} \ge (8.3\times10^{9} \mbox{ M}_{\odot}) \left(\frac{\nu f_{*}}{2\times10^{-3}}\right)^{3/2}
\left(\frac{1+z}{9}\right)^{-3/2}.
\label{rpm}
\end{equation}
{\c2 We plot the above equation, 
the minimum halo mass for pressure confinement for different values of $\nu f_{*}$
in Figure~\ref{rpcon}. The starburst
bubbles in the dwarf halos that we select for our study (filled diamonds: see \S4.1) 
are predicted to freely escape to intergalactic space if 
$f_{*}\ga0.1$ and to be completely confined by ram pressure of the infalling gas if $f_{*}\la0.001$
and $\nu\la0.5$. We discuss Figure~\ref{rpcon} further in \S 4.1. }
For a given $\nu f_{*}$, both the minimum halo mass for pressure
confinement in equation~(\ref{rpm}) and the halo mass with a fixed
virial temperature $T_{v}$ or circular velocity $v_{c}$ have the same redshift dependence,
$M_h \propto (1+z)^{-3/2}$.  Therefore, we can select a group of halos in
which the ram pressure of the infall can confine the bubbles and so
suppress the feedback efficiencies by a virial temperature or circular velocity 
cut-off: $T^{*}_{v}$ or $v^{*}_{c}$. We can not expect high feedback efficiencies
in dwarf galaxies with 
\begin{eqnarray} 
T_{v}\ga T^{*}_{v}\simeq (1.0\times10^{5} \mbox{ K}) \left(\frac{\nu f_{*}}{1\times10^{-3}}\right),
 \nonumber \\
v_{c}\ga v^{*}_{c} \simeq (52 \mbox{ km s}^{-1}) \left(\frac{\nu f_{*}}{1\times10^{-3}}\right)^{1/2}.
\label{tconf}
\end{eqnarray}


Without ram pressure from cosmological infall, a bubble will stall
when its expansion is balanced by the thermal pressure of the IGM.  If
we follow the same procedure as above, we find that the bubble will
only stall when its radius 
\begin{equation}
R_s \ga (10 R_v)  \left(\frac{\nu f_{*}}{1\times10^{-3}}\right)
\left(\frac{T_{IGM}}{2\times10^{4}\mbox{ K}}\right)^{-1},
\end{equation}
where we have assumed an ionized IGM.  
%
%
%

The presence of ram pressure does generally appear to be important to
the evolution of bubbles, and hence must be included when we compute
the metal, mass, and energy feedback efficiencies in numerical
simulations.

\section{Numerical Methods}
\label{sec:numer}

In this section, we outline the numerical methods we use to model
feedback from dwarf galaxies.  We first describe the model that we use
for the evolution of the background halo, then the implementation of
our models of starbursts.  Finally we examine an important detail of
our model: the performance of the tracer field that we use to trace
hot, metal-rich material from bubble interiors.

\subsection{Spherical Halo Model}
\label{sub:halo}

We model evolving dark matter halo potentials and the cosmological
infall of gas onto them using a method based on the one-dimensional
hydrodynamic computations of Meiksin (1994), which were developed to
study the structure and evolution of Lyman $\alpha$ clouds
gravitationally confined by dark matter minihalos. The infall code
described by Meiksin (1994) included standard CDM cosmology; we
modified it for a flat, $\Lambda$CDM cosmology.  We neglect the effects
of the repulsive force exerted by dark energy as it is negligible at
redshift $z\ga2$.  This can be seen by considering the expansion
parameter
\begin{equation}
H(t)^{2}=(\dot{a(t)}/a(t))^{2}=8\pi G\rho_{M}/3+\Lambda/3
\end{equation}
where $\Lambda$ is the cosmological constant and $\rho_{M} =
(1.88\times10^{-29}h^2$~g~cm$^{-3}) \Omega_0 (1+z)^{3}$. As $z$
increases, matter dominates.

The infall code follows the evolution of spherical shells of dark
matter based on a model of the spherical collapse of a Gaussian
density perturbation (Bond et al.\ 1988).  However, the dark matter
shells are not allowed to interpenetrate as they
%
collapse.
Instead each shell is brought to rest at its virial radius, building
a halo that reflects the core structure of the initial perturbation,
so the dark matter halo profile will not perfectly reproduce the results
of $N-$body simulations, especially in the central regions.
%
However, the central
regions of the halo are less important anyway, because the disk of the
galaxy dominates the potential in that region.  As we discuss in
\S\ref{sec:disk}, we therefore include a separate disk term in the
potential.  The halo potential is more important for determining the
density profile of the infalling gas at its edges, as well as
capturing the details of the late evolution of the bubble as it climbs
out of the potential well.

In the code, a linear Gaussian perturbation is evolved starting at
$z=200$.  For models that collapse at $z>5$, we assume that the IGM is
fully neutral with initial temperature $T_{bg}=550$~K determined by
the cosmic background radiation. For models that collapse at $z\le5$,
we assume that H is ionized and He is singly ionized, with initial
temperature $T_{bg}=10^{4}$~K, and we set the minimum temperature to
which the gas is allowed to cool to be $T_{min}=10^{4}$~K. This very
roughly captures the effects of UV background radiation after
reionization, which actually heats the IGM to $\sim2\times10^{4}$ K
(Sargent et al.\ 1980).  

{We assumed that the reionization was completed at $z_{ion}\sim
6$, and that the IGM was mostly neutral before $z_{ion}$, 
based on the onset of Gunn-Peterson trough against background quasars 
(Djorgovski et al.\ 2001; Becker et al.\ 2001). However, 
the recent measurement of a large electron scattering optical depth by 
{\it WMAP} implies that a significant fraction of the IGM was already 
ionized by $z\sim17\pm5$ (Kogut et al.\ 2003; Spergel et al.\ 2003).
If our models at $z>5$ are evolved in the reheated IGM instead, most 
of the gas ends up being pressure supported in 
halos with $T_v\approx5\times10^{4}$ K and $v_c\approx30$ km s$^{-1}$.
However, the differences in density and pressure 
distributions appear only 
in central regions of the halos, 
where the gas cools to form disks. 
We discuss the effects of suppressed cooling in \S4.2. }

The fluid equations are solved for baryons
that cool by optically thin radiative recombination losses from H and
He, dielectric recombination losses from He, bremsstrahlung, and
Compton cooling from scattering off the microwave background.  Cooling
is turned off in the central region where the potential is anyway
unphysical to avoid thermal instability. 
Collisional processes were included by Meiksin (1994) but were found
to be unimportant outside of the accretion shock, so they have been
neglected in our model to increase computational speed.  We also
neglect molecular hydrogen cooling.

The infall code produces a gas density distribution $\rho\propto
r^{-\alpha}$ with $\alpha\approx 2$--2.3 for $r \la R_v$
and $\alpha\la2$ outside, as predicted by the secondary infall
model by Bertschinger (1985).  
The density distribution at the center is not correctly modeled by the
infall code. However, 
we replace the central gas distribution with a rotationally
supported disk, as we describe in the next section.  The lack of
cooling in the central region does
produce shocks that convert some infall energy to thermal energy, 
and so the distribution of thermal or
ram pressure alone is not well represented physically in our infall
solutions, but the total pressure remains accurate.  
The details of
the energy and velocity profiles ultimately depend on how gas is
removed by instabilities that make molecular clouds and stars, which
we are still far from modeling.

Although we neglect the large-scale filamentary web predicted by
three-dimensional hydrodynamic simulations, the spherical infall code
is probably not too bad for the scale of our interest, because
virialized systems form at the intersection of the filaments, where
the accretion is more spherical within a scale of order the virial
radius $R_v$, closely corresponding to minihalos (e.g.\ Miralda-Escude
et al.\ 1996; Zhang et al.\ 1998).  Adaptive mesh refinement
simulations of first star formation by Abel et al.\ (1998; 2000; 2002) do
show that gas accretion occurs along filaments, producing a highly
structured, aspherical accretion shock even close to $R_v$.  The
results of our models in an idealized, spherical background can be
interpreted as a {\em lower limit} to the effectiveness of metal, mass, 
and energy feedback,
however, since a more dynamical, filamentary shape will always make it
easier for galactic winds to find escape routes from dense regions.

\subsection{Starburst Models}
\label{sub:2D}

Our models of starburst blowout generally follow MF99 (see their \S~4)
and Fujita et al.\ (2003: see their \S~4), so we briefly summarize our
methods below.

We compute the evolution of the blast wave from starburst supernovae
with ZEUS, an Eulerian, finite-difference, astrophysical gas
dynamics code (Stone \& Norman 1992; Clarke 1994), that uses
second-order van Leer (1977) advection. We use a quadratic artificial
viscosity to resolve shock fronts, and a linear artificial
viscosity to prevent instabilities due to strong shocks in standing
flows.  The former is necessary in any method that does not solve the
Riemann problem, while the latter only acts at the zone-to-zone scale,
and so does not affect larger-scale features of the solutions.  We use
the loop-level parallelized version ZEUS-3D, in its two-dimensional
form.  Runs were done on Silicon Graphics Origin 2000 machines using
eight processors, and typically took several days to a week, depending
on the parameters of the galaxy.

We assume azimuthal symmetry around the rotational axis of the galaxy,
and use a ratioed grid in both the radial and vertical directions to
capture the multiple length scales of the problem.  Our central,
high-resolution grid extends to $\sim 0.5 R_v$, and has typical zone
sizes of a few parsecs at high redshift to a few tens of parsecs at
lower redshift.  Zone size increases by a factor of ten across our
ratioed grid but in the outer parts of the grid still remains under a
few tens to hundreds of parsecs respectively.  The grids always
capture several $R_v$.  The actual values for the different models are
listed in Table~\ref{scale}.  The central resolution is high enough to
resolve the stratified atmosphere of disks with more than 50 zones.
This is found to yield converged results on the acceleration of
bubbles, and is less than or at least equal to the typical size of a
giant molecular cloud ($\sim10$--40~pc).
We use reflecting boundary conditions along the symmetry axis and
along the galaxy midplane and infall boundary conditions on the other
two axes. The infall boundaries are fed by interpolation from
intermediate results of the spherical collapse model described in the
previous section.  

To drive a constant luminosity wind, we add mass
and energy to a source region with a radius of five zones and a volume
$V$.  We choose the input rates based on the analytic solution by
Weaver et al.\ (1977) for a constant density medium with the density
$\rho_{0}$ and energy $e_{0}$ of the center of the galactic disk. 
The energy density
in each cell of the source zone is increased at a rate
$\dot{e}=L_m/V$, where $L_m$ is the mechanical luminosity from the
starburst.  The mass density is then increased at a rate
$\dot{\rho}=\rho_0\dot{e} / \eta e_0$ to roughly keep the specific
energy of the source region constant.

The amount of mass in this wind accounts not only for supernova
ejecta, but also mass that would have evaporated off the shells if we
had included thermal conduction in our model.  The evaporated mass
does, in fact, dominate the mass within a realistic bubble (Weaver et
al.\ 1977).  Therefore, we add a larger amount of mass by choosing
$\eta(L_m)$ for a given $L_m$, to reproduce the mass evaporation rate
predicted by Weaver et al.\ (1977) for a spherical bubble.  In this
way, we keep the density and temperature within the superbubble fairly
close to the analytic solution. See Fujita et al.\ (2003) for more
details.  With our choice of $\eta(L_m)$, the resulting interior
temperature is always within a factor of two of the expected
temperature.

We use a cooling curve for primordial gas in collisional ionization
equilibrium (Sutherland \& Dopita 1993) and also implement inverse
Compton cooling appropriate for the high-redshift universe.  
We employ 
a sharp cutoff in the cooling at
$T=10^{4}$ K.
This
not only distinguishes ionized gas which
cools, from neutral gas which does not cool at high redshift, but also
mimics the effect of UV background radiation at intermediate redshift,
in the absence of explicit models for photoionization and collisional
ionization in our code. In the high-redshift, dense universe, we
expect bubble shell shocks to be nearly isothermal due to efficient
collisional cooling ($\Lambda_{line}\propto n^{2}$), so the
collisional ionization equilibrium cooling curve is a reasonable approximation.
Strictly speaking, the assumption breaks down at the shock fronts,
where underionization will increase the cooling rate, and in the
evaporative flow off the shell, where overionization will decrease the
cooling rate.  Using the equilibrium curve overestimates cooling in
the interiors, ensuring that our feedback efficiencies are lower
limits to the true value.

The only place in the evolution of a bubble where substantial cooling
occurs is in the dense, swept-up shell, where collisionally-excited
radiative cooling determines the density.  The cooling in the shell
does not directly affect the dynamical evolution of a bubble, which is
controlled by the hot interior.  The cooling timescale of the interior
greatly exceeds the dynamical timescale of any realistic starburst
bubble (Mac Low \& McCray 1988). Instead, conductive and mechanical
energy fluxes
%
%
dominate
%
its energy budget.  
Rayleigh-Taylor instability occurs when the shell begins to accelerate
during blowout, so that the dense swept-up shell of ISM and halo gas is
supported by the low-density, hot gas of thermalized supernova ejecta
and evaporated shell gas inside.

The cooling curve is implemented in the energy equation in ZEUS-3D
with a semi-implicit method.  To solve the implicit energy equation, a
Newton-Raphson root finder is used, supplemented by a binary search
algorithm for occasional zones where the tabular nature of the cooling
curve prevents the Newton-Raphson algorithm from converging. The
empirical heating function from stellar energy input used in MF99 is
turned off in our study, Instead, the sharp cut-off in the cooling
function at $T=10^{4}$ K prevents the background atmosphere from
spontaneously cooling.

\subsection{Tracer Field and Its Performance}
\label{sub:tracer}

We use a tracer field $c$ both to follow metal enriched gas and to turn
off radiative cooling in the hot bubble interior, in order to prevent
mass numerically diffused off the dense shell from spuriously cooling
the interior.  The radiative cooling time of the interior is much
longer than the dynamical time of the bubble for the galaxies of our
study, so interior cooling is physically unimportant to the dynamics
of the bubbles (Mac Low \& McCray 1988). 
We set the cooling to be linearly proportional to the value of the tracer
field, with the metal-enriched, interior gas injected in the source region 
having $c=0$; hereafter we designate this as ``colored'' gas.   
We do not turn off Compton
cooling in the bubble interior because it is only linearly dependent
on the density.  
We advect a function 
\begin{equation}
    f(c)=\tan[0.99\pi(c-0.5)]
\label{etracer}
\end{equation}
of the tracer field with exactly the same algorithm as the density,
rather than the tracer field $c$ itself, so that $c$ will show a sharp
transition at the interface, even if $f(c)$ becomes quite smooth as
the result of numerical diffusion (Yabe \& Xiao 1993; MF99).

MF99 used the tracer field successfully to model modern, low-density
dwarf galaxies, but we find that the bubbles in our high-redshift
galaxies are severely poisoned by numerical diffusion, despite the
sharpening of the edges by the tangent transform.  Numerical diffusion
occurs due to the failure in advection over a large
density jump across a small number of zones.  

The cooling time of the shells formed by initially adiabatic shocks
is $t_{c} = 3kT_{s}/4n_0\Lambda(T_s)$, with postshock electron number density
$4n_0$ and shocked temperature $T_{s}=(3\mu_{s}/16k)v^{2}_{shock}$, 
where $\mu_{s}$ is the mass per particle in the ionized ISM, 
and $v_{shock}$ is shock velocity given by a similarity solution of a bubble
in a uniform density medium $n_0$ by Weaver et al.\ (1977).
We use the analytic estimate of the Gaetz and Salpeter (1983)
cooling function between $10^{5}$ and $10^{7}$ K from Mac Low \& McCray (1988), 
\begin{equation}
    \Lambda(T)=(1.0\times10^{-22}\mbox{~erg~cm}^{3}\mbox{ s}^{-1}) 
         \left(\frac{T}{10^6\mbox{~K}}\right)^{-0.7}
         \left(\frac{Z}{Z_{\odot}}\right).
\label{coolingcurve}
\end{equation}
This approximation agrees with the primordial cooling function of Sutherland \& Dopita (1993) 
that we use for our simulations within a factor of a few in the temperature range.
Then, we find the instantaneous cooling time of the shocked ambient gas 
to be very short in the early, high-density universe;
\begin{equation}
\tau_{sh}=(0.19~\mbox{ Myr}) L^{0.29}_{40} n^{-0.71}_{0} 
(Z/0.1 Z_{\odot})^{-1} \propto
(1+z)^{-2.13}
\label{coolsh}
\end{equation}
for a uniform density medium $n_{0}\propto(1+z)^3$,
so we expect the formation of
dense shells with density $n\propto n_0 {\cal M}^{2}$ where ${\cal M}$ is the Mach
number. 

We show the results of simple test simulations of the
performance of the tracer field in Figure~\ref{ftracer}.  Our standard
model
for this test
has fixed background density 
chosen at $z=13$
$
\rho_{bg} = 200\rho_{0}
=2.5\times10^{-25}$~g~cm$^{-3}$, which is an average
halo gas density at 
that redshift,
mechanical luminosity $L_m=10^{40}$
erg s$^{-1}$, and uses the primordial radiative cooling function by
Sutherland \& Dopita (1993).  The zone size is 4 pc.  We turned off
Compton cooling for this test because it unnecessarily slows down the computation,
but the degree of numerical diffusion that we find only becomes worse
with the inclusion of Compton cooling.

With the standard model, the tracer field already fails to pick up
25\% of initially marked interior gas by $t=5$~Myr after the onset of
starburst (at time step $\sim5000$).  The same tracer field performs
perfectly if the cooling is completely turned off (the adiabatic
model: C0), while it does worse if the cooling strength is linearly
increased by a factor of 100 (the extra cooling model: C100).  The C0
model does so well because the adiabatic shell is fully resolved over
$\sim20$ cells with density $\rho = 4\times 200\rho_{bg}$, while the
isothermal shells are unresolved, with density $\rho \la 200\rho_{bg}
{\cal M}^{2}$.  As the cooling strength increases, the peak density
approaches its theoretical prediction, and the numerical diffusion
gets worse.
%
In the low-resolution simulation LR with 12~pc zones, the performance
of the tracer field is even worse due to the failure to resolve the
shell.  Numerical diffusion increases in our models as unresolved,
high-density, swept-up shells fragment after the bubbles blow out of
the disks, increasing the area of contact between high and low density
gas.

We find that the degree of numerical diffusion near unresolved shells
does not improve substantially by changing the advection scheme from
second-order, van Leer method to third-order Piecewise Parabolic
Advection (PPA) method (Clarke 1988; Colella \& Woodward 1984).  We
also find that changes in the size of source regions do not
influence the performance of the tracer field. 

The tracer field controls cooling, preventing it in gas marked as
interior ejecta. 
{\cc When hot gas is mixed into regions colored as
external, it is allowed to cool.}  As mass from the shell has usually
numerically diffused into this gas, it cools unphysically quickly, so as diffusion
proceeds, the interior energy of the bubble is artificially
dissipated.  Our goal in this study is to compute metal/mass ejection and
energy transport efficiencies by simulating the bubble dynamics in
high-redshift galaxies with the tracer field. However, our simulations
only yield strict lower limits on the efficiency of metal, 
mass, and energy feedback because
of the numerical diffusion.  To solve this problem we will eventually
need either high enough resolution to cleanly resolve shells, perhaps
using adaptive mesh refinement, or else use of a true contact tracer
as was done by Mac Low et al.\ (1989).

\section{Galaxy and Star Formation Parameters}

\label{sec:disk}

\subsection{Galaxies}
\label{sub:galaxies}

We model objects representing 1--3$\sigma$ peaks in the $\Lambda$CDM
density perturbation spectrum (see Fig.~\ref{selh}).  We focus on
second generation systems with virial temperature $T_{v}> 10^{4}$ K
that can cool through line cooling by H and He. We show the galaxies
we model with
%
halo masses 
%
$M_h=5\times10^{8}$--$5\times10^{10}$~M$_{\odot}$ that are
expected to collapse at $z=3, 5, 8,$ and 13 in Figure~\ref{selh}.  We
make a list of the properties of our model galaxies in
Table~\ref{scale}.

In these second generation objects, the cooling time is much shorter
than the dynamical time, so the gas will cool nearly isothermally at
$\sim10^{4}$ K and rapidly flow to the centers.  We assume that it settles in
tightly, bound, rotationally supported disks (Navarro \& White 1994)
without much fragmentation (Kashlinsky \& Rees 1983; Oh \& Haiman
2002).  Such disks are mostly gravitationally stable, but a large
fraction of out-of-equilibrium, residual, electrons expected during
initial line cooling drives H$_2$ formation through gas-phase
reactions catalyzed by H$^-$ (Mac Low \& Shull 1986; Shapiro \&
Kang 1987). Radiation from the H$_2$ allows the gas to cool further
and collapse (Oh \& Haiman 2002).  

Figure~\ref{selh} also shows the Jeans mass for gas in photoionization
equilibrium with the background UV radiation with
$T\approx2\times10^{4}$~K.
{\c2 We chose two galaxies with $M_h=5\times10^{8}$~M$_{\odot}$ and  
$v_{c}\approx30-40\mbox{ km s}^{-1}$ at $z=8$ and 13, assuming the effect of 
the UV background radiation to be negligible
at $z>6$. 
The cooling of gas in halos with $v_c\la30$ km s$^{-1}$ is suppressed 
in the reheated IGM (e.g. Bond, Szalay, \& Silk 1988; Meiksin 1994; Efstathiou 1992; Thoul \& Weinberg 1996). 
Therefore, we would 
not expect starbursts in such small halos at $z=8$, 
if they are found in the IGM fully ionized by that time. 
The results in these halos should be interpreted in that light. 
We also chose three galaxies with $M_h=5\times10^{9}$~M$_{\odot}$ and
$v_{c}\approx40-60 \mbox{ km s}^{-1}$ at $z=3$, 5, and 8.
These galaxies fall below the Jeans mass,
but 
hydrodynamic studies of dwarf halo formation show that $\sim50\%$
of baryons within the virial radius can still cool as long as
$v_{c}\ga50 \mbox{ km s}^{-1}$ 
(Vedel et al.\ 1994; Thoul \& Weinberg 1996; Navarro \& Steinmetz 1997).
We can still expect moderate 
starbursts
in these halos. }

As a comparison, we also plot in Figure~\ref{selh} the minimum halo mass for first
generation, or Population III, systems
with $T_{v}<10^{4}$ K that
can only cool through molecular hydrogen formation, as computed by
Tegmark et al.\ (1997) for a standard CDM model.  We do
not select any of these first generation objects because their star
formation efficiency is expected to be very low due to the presence of
internal and external photo-dissociating radiation (Omukai \& Nishi 1999; Glover \& Brand
2001; Haiman et al.\ 2000).  We are far from realistically computing
the efficiency of cooling and star formation even in second 
    generation
systems that we study, and cannot
definitively predict in which halos starbursts will occur at a
given redshift. Instead, we perform a parameter study that explores a
large range of dwarf galaxies to understand the physical processes of
stellar feedback in different environments.

In Figure~\ref{rpcon}, we plot our chosen galaxies on top of the
pressure confinement criterion given by equation~(\ref{rpm}) for 
different values of $\nu f_{*}$.
With $f_{*}=0.1$, equation~(\ref{rpm})
predicts that any bubble in a dwarf galaxy will escape to the IGM
unimpeded by the ram pressure of infalling gas, unless the bubble has
lost so much of its energy that 
{\cc $\nu<0.1$} and the galaxy formed
from an extremely large density perturbation (see Fig.\ \ref{selh}).
With $f_{*}=0.001$, 
{\cc most second generation systems with $T>5\times10^4$ K }are predicted to
be confined by ram pressure with $\nu\le0.5$.  The most interesting
case in Figure~\ref{rpcon} is with $f_{*}=0.01$.  
 If the fraction of energy preserved in the bubbles 
{\cc $\nu \le 0.2$}, the bubbles in halos
with $M_h\ga5\times10^{10}$~M$_{\odot}$ from $\sim2\sigma$ peaks at $z=3$--5 
will be confined by the ram pressure of the infall. If
$\nu\la 0.1$, even bubbles in smaller halos with
$M_h\approx5\times10^{9}$~M$_{\odot}$ will begin to be confined by the
ram pressure. 
However, we can expect the bubbles in halos with $M_h\la 10^{9}$~M$_{\odot}$ 
{\cc ($T<5\times10^4$ K) } to escape to the IGM, 
unless most of their energy is lost
to the surroundings. Figure~\ref{rpcon} suggests high metal, mass, and 
energy feedback efficiencies in small high-redshift dwarf galaxies with $f_*=0.01$ 
and larger galaxies with $f_*\gg0.01$.
In \S\ref{sec:results}, we compare the results of our simulations with
these predictions, and try to constrain $\nu$.

\subsection{Disks}
\label{sub:disks}

In our simulations, we set up a disk with a uniform surface density
profile, $\Sigma=M_{d}/\pi R^{2}_{*}$, where $R_{*}$ is the size of
the disk, as in MF99.  We choose 
a uniform surface density profile
for numerical simplicity, although an exponential surface density
distribution is observed in local blue compact dwarf galaxies (van Zee et
al.\ 1998). However, bubble evolution is mainly governed by the
vertical distribution of gas within the central region of the disk,
where a constant surface density is a good approximation.
 
To compute the size of a galaxy as a function of halo
mass and collapse redshift we follow Mo, Mao, \& White (1998), using
an isothermal halo profile.  The size of a disk with a uniform surface
density is
\begin{equation}
\label{Rd}
R_*=\frac{3}{\sqrt{2}}(\frac{j_{d}}{m_{d}})\lambda R_v,  
\end{equation}
where $R_v$ is the virial radius of a halo, 
\begin{equation}
R_v = (16 \mbox{ kpc}) (1+z)^{-1}
\left(\frac{M_{h}}{10^9 \mbox{ M}_{\odot}}\right)^{1/3}  
(\Omega_{0} h^{2})^{-1/3},
\label{rv}
\end{equation} 
where we use the same approximation for the $\Lambda$CDM model as in
\S\ref{sub:halo}, and we set the dimensionless spin parameter
$\lambda= 0.05$, approximately at the peak of its distribution (e.g.\
Cole \& Lacey 1996).  We assume that a fraction of halo mass,
$m_d=M_{\rm d}/M_{\rm h}$, settles into a non-self gravitating,
rotationally supported disk in a singular isothermal halo with a
fraction of halo angular momentum, $j_d=J_{\rm d}/J_{\rm h}$.  This
method provides a good fit to the observed size distribution of
galactic disks 
%
for
%
$j_d/ m_d = 1$ (Mo, Mao, \& White 1998).

The baryonic content of each halo
%
is
%
$M_{g} = (\Omega_{b}/\Omega_{0})M_{h}$, the amount of gas within the virial
radius $R_v$ if the gas simply follows the dark matter.  
We choose the mass which goes into the disk $M_{d}$ by subtracting the
result of the infall solution described in \S\ref{sub:halo} from
$M_g$.  Typically, $M_{d} \simeq 0.5 M_{g}$, but it varies by 10\% among
models with different cooling histories.  
{\c2 The disks containing $\sim50\%$ of cooled baryons are appropriate for 
dwarf galaxies with $v_{c}\ga50\mbox{ km s}^{-1}$, but not for those with
$v_{c}\approx30\mbox{ km s}^{-1}$, if the infall gas was initially reheated to 
a few $\times10^{4}$ K (see \S 4.1). 
In halos with $v_c\approx30$ km s$^{-1}$ at $z<z_{ion}$, 
most of the halo is pressure-supported, and
only small fractions of 
the gas mass 
$M_{g}$ 
can cool to form disk-like objects.
The evolution of bubbles in such galaxies will only differ within the 
virial radius $R_v$ from our standard models 
with disk masses as high as $M_d\approx0.5M_g$.
The qualitative results on metal, mass, 
and energy feedback efficiencies should not differ much, as long as 
disks or disk-like objects with stratified ISM 
are assumed to form,
because 
the bubbles blow out of them quickly to interact
with the halo gas and the IGM. However, star formation efficiency $f_*$ 
should be 
    strictly
$\la0.01$ in the halos with $v_{c}\approx30\mbox{ km s}^{-1}$
if the collapse redshift $z<z_{ion}$.  }

We initially set up the disk in hydrostatic equilibrium with a
potential field due to the sum of the dark matter halo at the collapse
redshift and the potential of a cold, thin disk representing earlier
stars formed in fragmented gas or in the disk
prior to any starburst.
The dark
matter potential computed by the infall code evolves as a function of
time, but we do not modify the initial potential within $R_v$ in our
two-dimensional models. Since our dark matter profile is not realistic
in the center, as discussed in \S\ref{sub:halo}, we must include the
cold, thin, disk potential to model a disk with an exponential
atmosphere that resembles the gas distribution observed in the Galaxy and
nearby galaxies (Lockman et al.\ 1986; Neeser et al.\ 2002).
To do so, we assume the 
    presence 
of rather massive, thin disks containing $0.03M_g$. 
This approximation of a fixed, cold disk potential is
probably reasonable, as the dark matter still dominates the potential
everywhere.

We assume that the gas in the disk is in the warm, neutral phase with
some turbulent contribution provided by winds and supernovae from the
same small-scale star formation that formed the cold, thin disk,
yielding an effective sound speed of $c_s\sim 10\mbox{ km s}^{-1}$
with corresponding temperature, $T\approx10^{3.7}$~K.  
Modeling 
this 
diffuse, warm medium is
important for the dynamics of bubbles, since the bubbles will
immediately blow out of the cold gas disks where star formation is
actually expected to occur (Mac Low \& McCray 1988).

\subsection{Starbursts}
\label{sub:starbursts}

At the beginning of our simulations, we set up an instantaneous
starburst at the center of each disk.  The expansion of a starburst
bubble can be considered to be an extension of superbubble dynamics in
an OB association in a disk galaxy. The stellar winds create a hot,
low-density cavity in the ISM, and repeated supernovae excavate a
larger hole as they sweep the gas into a thin, dense shell (Tomisaka
\& Ikeuchi 1986, Mac Low \& McCray 1988, Mac Low et al.\
1989). Stellar winds are weaker in low-metallicity stars (e.g.\
Kudritzki 2002), but the initial few SNe will still create a similar
hot cavity in high-redshift galaxies.  Once the hot cavity forms,
discrete supernova explosions generate blast waves that become
subsonic in the hot interior, and hence can be treated as a continuous
mechanical luminosity $L_m$ in the study of bubble dynamics (Mac Low
\& McCray 1988).

It is, of course, an approximation to assume that all the mass
available for the starburst turns into stars instantaneously and that
the starburst is concentrated in a single spot at the center of a
disk.  Local observations of dwarf starburst galaxies show that
multiple starburst clumps are scattered around the disks instead
(e.g.\ Vacca 1996; Martin 1998).  High-redshift disks are much smaller
than present-day dwarf galaxies, roughly $R_{*}\propto (1+z)^{-1}$, but the
smallest galaxy that we model extends over a few hundred pc, which is
an order of magnitude larger than a typical molecular cloud ($\sim10$
pc). 
Therefore, we expect the same or at least similar starbursts in
high-redshift galaxies, too, as long as cooling is efficient.  The
starburst clumps are made up of OB associations or clusters. Since the
largest OB associations or clusters have masses of a few million solar
masses, we could actually think of our starbursts as being powered by
a single star cluster as long as we do not assume 
too high a star formation efficiency in our model.  

In our study, we use uniform mechanical luminosities computed using the 
Starburst 99  model with $Z=0.001$ (Leitherer et al.\ 1999).
The Starburst 99 model is based on a power law initial mass function
which approximates the classical Salpeter (1955) function, with
exponent $\alpha=2.35$ between low-mass and high-mass cutoff masses of
$1$~M$_{\odot}$ and $100$~M$_{\odot}$ respectively.  The average
mechanical luminosity predicted by the Starburst 99 model is
$L_m=2.7\times10^{40}$~erg~s$^{-1}$ when $10^{6}$ M$_{\odot}$ of gas
is instantaneously converted to stars of mass between 1 M$_{\odot}$
and 100 M$_{\odot}$. {\c2 Note that approximately the same 
mass
of stars below 1 M$_{\odot}$ will 
form at the same time. } 
The Starburst 99 model is based on observations
of star-forming and starburst regions in the local universe.  However,
various low to intermediate-redshift observations suggest that the
initial mass function may have been top heavy in the early universe
(see the summary by Larson 1998), a suggestion supported by recent
studies of Population III stars (Abel et al.\ 2000; 2002; Bromm,
Coppi, \& Larson 1999; 2002; Nakamura \& Umemura 1999; 2001).  If the
initial mass function is indeed top-heavy at high redshift, the
fraction of O and B stars which contribute to the mechanical
luminosity increases, so the conversion of the same amount of gas to
stars can provide a higher mechanical luminosity than predicted by
the Starburst 99 model.  Within the uncertainties, we assume that the
mechanical luminosity of a burst is proportional to an amount of gas
%
%
$f_{*}M_{g}$ available for a starburst 
%
as follows: $L_m=2.7\times10^{40}
f_{*}(M_{g}/10^{6}$~M$_{\odot})$~erg~s$^{-1}$.

Our mechanical luminosities range from
$10^{39} < L_m < 10^{43} \mbox{ erg s}^{-1}$ produced when stars form
from 6--7$\times10^{4}$~M$_{\odot}$
of gas in the smallest halo with $f_{*}=0.001$ 
and 6--7$\times 10^{8}$~M$_{\odot}$ in the largest halo with $f_{*}=0.1$.  
These mechanical luminosities are equivalent to a supernova with
energy of $10^{51}\mbox{ erg}$ exploding every $\sim17,000$~yr to
every $\sim1.7$~yr, respectively. Energy input is continued for a
period of 50~Myr, roughly the lifetime of the smallest B star to go
supernova (McCray \& Kafatos 1987). 
{\cc Most cosmological simulations with supernova feedback
employ $f_*\ga 0.1$ to test its maximum effects.}

We consider starburst efficiencies $f_{*}$ in our galaxies of 0.001,
0.01, and 0.1.  Numerical simulations of first star formation suggest
$f_{*}\approx0.001$ in metal-free primordial clouds (Abel et al.\
2000; 2002; Bromm, Coppi, \& Larson 1999; 2002).  Oh \& Haiman (2002) estimate an
initial star formation efficiency by molecular hydrogen cooling alone
to still be $f_{*}=0.001$, even in second generation disks.  The star
formation efficiency in such disks is not well constrained because of
the presence of internal, photo-dissociating, UV radiation, but we can
expect metal cooling to enhance the star formation efficiency, if the
halo gas is pre-enriched at least to $Z=5\times10^{-4}$ Z$_{\odot}$ by
previous generations of stars (Bromm et al.\ 2001; Oh \& Haiman 2002; 
Wada \& Venkatesan 2003).
Therefore, the star formation could actually be rather efficient in
second generation objects.  We think that $f_{*}\approx
0.01$ may be a reasonable guess for the initial major starburst in a
galaxy. Starburst efficiencies as high as $f_{*}=0.1$ are probably
not realistic.  Since we are performing a parameter study, however, we
do consider them, in order to clearly delineate the starburst
efficiencies with which efficient metal/mass ejections and energy transport are
expected in a halo at a given redshift.

\section{Results}
\label{sec:results}

We have explored the effects of repeated supernovae with mechanical
luminosity linearly proportional to the amount of mass converted to
stars, $f_{*}M_{g}$, in seven examples of early dwarf galaxies.
Figure~\ref{z8g95} shows time histories of the density distributions
of our model with $M_h=5\times10^{9}$ M$_{\odot}$ at $z=8$, with
starburst efficiencies $f_{*}$ ranging from 0.001--0.1.  
The bubbles
shown in Figure~\ref{z8g95} are representative of our other models in
a number of ways as we now describe. (Figures~\ref{figures}{\em a--c}
show the density distributions of all of our models at $t=100$ Myr.) 

The bubbles quickly begin to accelerate vertically through the
stratified ISM of the galaxies, and blow out of the disks by $t\sim
10$~Myr after the onset of the starbursts.  With sufficiently high
central mechanical luminosity ($f_{*}=0.01$--0.1), the bubbles sweep up
the entire disks by $t=30$~Myr.  This behavior occurs in all of our
galaxies, because our high-redshift disks are very small, less than
$\sim1$ kpc in size, and are highly stratified in their deep potential
wells.  Most of the ISM remains bound, so these disks are not yet
``blown-away'' in the sense used by De Young \& Heckman (1994) and
MF99.  In fact, we will show below that most of the ISM remains bound
to the halo potentials if $f_{*}\la0.01$.

The internal termination shocks outside the freely expanding winds and
ejecta are clearly seen at $t=10$--30 Myr, surrounded by hot, pressurized
regions of shocked winds and supernova ejecta.  The hot, pressurized
regions drive shocks into the surrounding ISM and halo gas, sweeping
it up.  In these high-redshift, dense galaxies, gas behind the outer
shocks tends to cool efficiently, so that the shocks are usually
isothermal. Outside the galactic disks, the shocks are no longer isothermal.
Figure~\ref{tempden} shows the formation of dense shells, 
cooled from $T_{shock}\approx10^{5}$ K, 
 driven by the hot, pressurized gas at $T\approx10^{7}$ K behind, 
as the bubble blows
out of the disk at $t=10$ Myr, with $f_*=0.01$ (middle panel in Figure~\ref{z8g95}). 
The swept-up gas forms dense, cold, thin shells except in
the cases with $f_{*}=0.1$, when the shocks are so fast that cooling
remains inefficient.  

Shell formation is extremely sensitive to the resolution employed, and
is not fully resolved in our simulations (see Figure~\ref{tempden} and
also the resolution study of
shell formation in Fujita et al.\ 2003). Resolving the shell is not
important to following the overall dynamical evolution of bubbles
driven by the thermal energy of the hot pressurized regions (Castor,
McCray, \& Weaver 1975; Weaver et al.\ 1977), but is important to
follow the details of shell fragmentation due to Rayleigh-Taylor
instability when the bubbles accelerate.
Figure~\ref{z8g95} shows examples of shell fragmentation.  The
development of
the Rayleigh-Taylor instability in our model is limited not
only by the resolution, but also by the assumption of azimuthal
symmetry.
{\cc 
Mac Low et al.\ (1989) pointed out that in an axisymmetric blowout,
the typical spike and bubble structure of the Rayleigh-Taylor
instability is limited to rings.  Thus the detailed structure of the
fragments will be different in 3D.}
However, the instabilities seen in our simulations effectively allow
hot gas to accelerate beyond the cold, swept-up shells, which is the
qualitative result of interest.  Figure~\ref{z8g95} shows that for
$f_{*}\ge0.01$ the hot, metal-enriched gas expands into the IGM well
beyond the virial radius of the halo, leaving the fragmented shells
behind, which eventually fall back into the center. The hot gas is
confined in the vicinity of the disk if $f_{*}=0.001$. Similar behavior
can be seen in Figure~\ref{figures}{\em a--c} for all of our model
galaxies.

We now discuss our quantitative results.  We compute metal and gas ejection 
and energy transport efficiencies in \S\ref{sub:feedback}; discuss the
effects of external ram pressure from infalling gas on the
evolution of the bubbles and so on the feedback efficiencies in
\S\ref{sub:press}; and estimate in \S\ref{sub:time} how long it takes
for swept-up but still bound mass to fall back to the center of a
halo, where it will be available to fuel the next starburst.

\subsection{Feedback Parameters}
\label{sub:feedback}

In order to quantify feedback, we measure metal ejection 
efficiency $\xi_{met}$, mass ejection efficiency $\xi$, 
and energy transport efficiency $\zeta$. 
In each case, we measure the content of gas that has been accelerated
to outward velocities higher than the local escape velocity defined by
the gravitational potential. Because of the limitations of our
model, we actually measure the efficiencies 
when the bubbles reach 2$R_v$, where we assume the low-density IGM 
to begin.  This choice is rather arbitrary, and is only motivated by
the idea that the bubble should be well outside the halo virial radius
$R_v$, in order to see the effect of the ram pressure of infalling gas.

Recall that we use a one-dimensional, spherical background potential
(\S\ref{sub:halo}).
%
%
Simulations show filamentary structures are anchored in spheroidal
halos (Zhang et al. 1998). Thus the spheroidal approximation is a
good one within a virial radius, but becomes progressively worse
beyond. In reality, bubbles such as those we model would possibly
find their way out to low-density regions between the filaments.
%
When the bubble finds such a funnel for escape, we
expect that any gas carrying supernova energy and metals that has
velocity greater than the escape velocity of the halo potential
will escape nearly unimpeded into the IGM. 

\subsubsection{Metal Ejection Efficiency}
\label{subsub:metal}

The metal ejection efficiency was defined by MF99 as
\begin{equation}
\xi_{met, MF}=M_{c,esc} / M_{SN}, 
\label{memf}
\end{equation}
where $M_{c,esc}$ is the 
mass of metal-enriched gas moving with speed greater than the local escape velocity
and colored by tracer field with $c<0.6$; and $M_{SN}$ is the total mass injected at the center
with $c=0$ initially, 
which accounts not only for winds and supernova ejecta, but also for the mass
evaporated off the swept-up shell by thermal conduction. The colored mass 
$M_c$ traced is not sensitive to the choice of the cut-off number (0.6) in 
the absence of numerical diffusion. The mass
$M_{SN}$ is found inside the hot, pressurized, region in which metals
are assumed to be mixed well.  Equation~(\ref{memf}) is based on the
assumption that the injected mass $M_{SN}$ is preserved at the time of
its measurement, without being mixed into the ambient medium
physically or numerically.  However, the numerical diffusion severely
poisons the colored gas as discussed in \S\ref{sub:tracer}, and our
tracer field only picks up $\sim60\%$ of originally colored gas within
5--10~Myr after the onset of the starbursts.

At the same time, we observe a significant amount of physical mixing
between the metal-enriched, colored gas and the ambient 
gas. We show in Figure~\ref{metalmix} an example of metal mixing
inside a bubble in a halo with $M_h=5\times10^{8}$~M$_{\odot}$ at
$z=8$. The dense, swept-up shell of ambient gas fragments due to
Rayleigh-Taylor instability when the bubble blows out at $t\approx10$
Myr.  Figure~\ref{metalmix} shows the density distribution of the
bubble at $t=19$ Myr, and shows that the fragmented shells do mix with
the hot interior gas very effectively. This physical mixing of gas
accelerates numerical diffusion further.  Unfortunately, we have no
systematic way of tracking physical metal mixing, nor of
distinguishing it from numerical diffusion, but it is important to
quantify the fraction of the escaping gas enriched with metals to any
extent.  

Therefore, it is not appropriate for us to use equation~(\ref{memf}),
since we severely underestimate the metal ejection efficiencies when
we can not track down the total amount of originally colored
mass. Instead we define the metal ejection efficiency as
\begin{equation}
\xi_{met}=M_{c, esc}/M_{c}, 
\label{me2}
\end{equation} 
where now the color cut-off, $c < 1-\epsilon$, with $\epsilon$ chosen
small. In this way, we are sure to pick up any metal-enriched gas 
(that is, gas that was injected with $c=0$),
and still obtain a strong lower limit to the actual metal ejection
rate, since $M_{c}\gg M_{SN}$.   
{\cc Figure~\ref{cmass} shows the colored mass $M_c$ as a function of 
$\epsilon$ for the halo with $M_h=5\times10^8$ M$_{\odot}$ at $z=8$ for
star formation efficiencies $f_*=0.001$, 0.01, and 0.1. The masses $M_c$
were measured when the bubbles reached $2R_v$ at $t=100$ Myr when $f_*=0.001$ 
and 0.01 and at $t=50$ Myr when $f_*=0.1$. The mixing between the 
colored gas and the ambient gas proceeded far by $t=100$ Myr 
for the bubbles with $f_*=0.001$ and 0.01 (see Figure~\ref{figures}{\em a}), 
so the values of $M_c$ are insensitive to the choices of $\epsilon$ 
as long as they are very small. 
The asympotic value of $M_c$ as $\epsilon$ decreases is not reached for the 
case with $f_*=0.1$ because it was measured at an earlier time. However, 
the bubble is so powerful as to reach $2R_v$ by that time, so any values of 
$M_c$ will 
yield 
$\xi_{met}\approx1$ with $10^{-4}\le\epsilon\le0.6$. }
We choose $\epsilon\approx10^{-4}$--$10^{-3}$ to enforce $M_{c} < M_g$, the
total mass of baryons in the halo.

The metal ejection efficiencies $\xi_{met}$ are nearly zero in all the
halos that we study when the star formation efficiency $f_{*}=0.001$
and close to 1 with $f_{*}=0.1$ (Table~\ref{result}). These
results are consistent with our prediction in \S\ref{sec:predict} that
in the presence of infall ram pressure, the bubbles in second
generation objects can escape to the IGM if $f_{*}=0.1$, but will
be confined within the halo potentials if $f_{*}=0.001$ and
$\nu\la0.5$.  We find that the metal ejection efficiencies are largely
determined by the value of $f_{*}$ that we assume.  Recall that
$f_{*}=0.1$ is likely to be unrealistic for instantaneous starburst in 
second generation systems.

The most interesting results are therefore the values of $\xi_{met}$
with $f_{*}=0.01$.
Metal ejection efficiencies are
low in 
halos with
$M_h=5\times10^{10}$~M$_{\odot}$ at $z=3$ and 5 and 
even
with $M_h=5\times10^{9}$~M$_{\odot}$ at $z=8$.  However, high metal ejection
efficiencies $\xi_{met}\ga 0.5$ occur in halos with virial
temperatures $T_{v}\la1\times10^{5}$~K or circular 
velocities $v_{c}\la50\mbox{ km s}^{-1}$ at all redshifts we
studied. These galaxies may make significant contributions to the
metal-enrichment of the IGM.  
On the other hand, larger galaxies with $v_c\approx100$ km s$^{-1}$ 
allow efficient cooling, so
star formation efficiencies $f_*\gg0.01$ may be plausible.
These galaxies could then
produce and expel more metals to the IGM.

If we assume that each supernova produces $\sim 3$~M$_{\odot}$ of
heavy elements (Ferrara \& Tolstoy 2000), a halo with 
dynamical mass $M_h$ can expel metals with mass
\begin{equation} 
M_{met}\approx (8\times10^{4}\mbox{ M}_{\odot})
\left(\frac{\xi_{met}}{0.5}\right) \left(\frac{f_{*}}{0.01}\right)
\left(\frac{M_h}{5\times10^{8}\mbox{ M}_{\odot}}\right). 
\label{SNmetal}
\end{equation}
If we further assume that those metals mix uniformly, we can estimate the average
metallicity of the universe from halos at a given redshift $z$; 
\begin{equation}
\bar{Z} \simeq (6 \times 10^{-4} \mbox{ Z}_{\odot})\int^{M_{max}}_{M_{min}(z)} 
\left(\frac{N_{h}(M_h,z)dM_h}{1~\mbox{Mpc}^{-3}}\right) 
\left(\frac{\xi_{met}(M_h,z)}{0.5}\right)
\left(\frac{f_{*}(M_h,z)}{0.01}\right) \left(\frac{M_h}{5\times10^{8} \mbox{M}_\odot}\right),
\label{metalest}
\end{equation}
where $N_h(M_h,z) dM_h$ is the comoving number density of halos at $z$ with masses
between $M_h$ and $M_h+dM_h$,  
predicted from Press-Schechter theory (Press \& Schechter 1974, Bond et al.\ 1991) 
in our $\Lambda$CDM model, and $M_{max}$ and $M_{min}(z)$ are the maximum and minimum 
halo masses to be incorporated into the calculation. We set $M_{min}(z)=5.8\times10^{8}$ M$_{\odot}
\left[(1+z)/9\right]^{-3/2}$, corresponding to $v_c=30$ km s$^{-1}$.
Although the metals may well be distributed highly inhomogeneously, equation~(\ref{metalest}) provides 
a lower limit to the average metallicity of intergalactic clouds where the expelled metals can 
reach. 

Assuming $\xi_{met}f_*$ to be constant, we plot the average metallicity $\bar{Z}
\propto \xi_{met}f_*$ (eq.~\ref{metalest}) as a function of mass $M_{max}$
at $z=3$, 5, 8, and 13  in Figure~\ref{Zmetal}.  
We chose $\xi_{met}f_*=0.005$, which means for example, $f_*=0.01$
and $\xi_{met}=0.5$ for halos with $v_c\approx50$ km s$^{-1}$ and 
$f_*=0.05$ and $\xi_{met}=0.1$ for halos with $v_c\approx100$ km s$^{-1}$.
Although we neglect the dependence of $\xi_{met}$ on halo properties for a given $f_*$, 
Figure~\ref{Zmetal} shows a qualitative trend of metal contributions
from halos at different redshifts. 

In Figure~\ref{Zmetal}, the metal contribution seems to be poor from halos at $z=13$, but
rather substantial from halos at $z=8$. The average metallicity
at $z=8$ from halos with $v_c\la50$ km s$^{-1}$ alone is already 
approaching $\bar{Z}\approx10^{-3}$ Z$_{\odot}$, the lower limit of the metallicity
observed in low-density Lyman $\alpha$ clouds at $z\approx 3$--5.5. 
Such halos correspond to $\sim2\sigma$ peaks in our $\Lambda$CDM
primordial density perturbations. 
The early enrichment at $z\approx8$ may explain the absence of kinematic disturbance observed
in low-density Lyman $\alpha$ clouds at $z\approx3$ (Rauch et al.\
2001b; Madau, Ferrara, \& Rees 2001),
by which time the kinematic effects of the wind may have dissipated
and the gas resettled onto the filaments.
Note however, the metal contributions from halos $v_c\la50$ km s$^{-1}$ are a factor of
a few larger at lower redshifts, since they are more abundant (1--2$\sigma$ peaks) with larger
halo masses, and so can produce more metals with the same $\xi_{met}f_*$. 
Figure~\ref{Zmetal} also shows that the larger galaxies with $v_c\approx100$ km s$^{-1}$
begin to make significant contributions, yielding $\bar{Z}\ga10^{-2}$ 
Z$_{\odot}$ at $z\approx3$--5. The high H{\sc i} column density Ly$\alpha$ clouds 
($\log N_{\rm \mbox{H{\sc i}}}>14.4$) with $Z\approx10^{-2}$ Z$_{\odot}$ (Cowie et al.\ 1995) 
might have been recently enriched by such dwarf galaxies then. 
This picture of recent enrichment is also supported by 
large turbulent motions ($v_{rms}\approx70$ km s $^{-1}$) 
observed in high column density C{\sc iv} systems (Rauch et al.\ 2001b).

Our estimates in Figure~\ref{Zmetal} based on our numerical simulations 
suggest that both dwarf starburst galaxies with $v_c\approx30$--50 km s$^{-1}$ 
and with $v_c\approx100$ km s$^{-1}$
can be very efficient in spreading metals, 
but to different regions of the IGM (low and high density clouds) at different times 
($z\approx8$ and $z\la5$).
We find that 
only galaxies collapsing from 1--$2\sigma$ peaks are 
abundant
enough to make significant contributions to the metal enrichment of the IGM. 

Metal ejection efficiencies $\xi_{m,MF} \sim 1$ were found by MF99
in all of the local dwarf galaxies that they modeled, even with star
formation efficiency $f_{*}=0.001$, while we found negligible metal
feedback with $f_{*}=0.001$ in high-redshift galaxies. We showed in
\S\ref{sub:press} that the difference between our results and those of
MF99 is not the higher densities $\rho\propto (1+z)^3$, because
bubbles quickly blow out of the smaller disks with smaller scale
heights expected at higher redshift.
Rather, it is the ram pressure from cosmological infall
that makes the difference.

\subsubsection{Gas Mass Ejection Efficiency}
\label{subsub:mass}

We define the efficiency of gas mass ejection as
\begin{equation}  
\xi = \frac{M_{esc}}{(\Omega_{b}/\Omega_{0}) M_{h}}, 
\label{mass}
\end{equation}
where $M_{esc}$ is the mass moving with velocity $v\ge v_{esc}$, and
the denominator is the mass of baryons initially within $R_v$,
%
%
as we assumed they initially simply follow the dark matter in our models.
%
We list $\xi$ for each of our models in Table~\ref{result}.

The mass $M_{c,esc}$ of escaping, metal-enriched gas is a negligible
fraction of $M_{esc}$.  The bulk of the mass escaping is some
combination of ISM and infalling halo gas.  Since we do not
systematically track the disk gas in our simulations, we identify gas
with angular momentum as disk gas, although there is some transfer of
angular momentum between the disk and ambient gas due to numerical
diffusion and physical mixing.  Our accuracy is limited, so we can
only observe whether $\ga50\%$ of
the ISM is moving faster than the escape
velocity or not.  
Halos in which this occurs 
are marked 
as having had their ISM blown away in Table~\ref{result}.

Table~\ref{result} shows that the mass ejection efficiencies 
$\xi \ga 1$
with $f_{*}=0.1$ and are nearly zero with
$f_{*}=0.001$, consistent with the metal ejection efficiencies.
More than half of the disk gas escapes the
potentials when $f_{*}=0.1$ in all the halos except one.
Blow away is possible with $f_{*}=0.1$ because high-redshift galaxies
are very small and easily swept away by powerful bubbles; however, our
result should be interpreted carefully because of our assumption of an
instantaneous starburst at a single site (see the discussion in
\S\ref{sub:starbursts}).  With $f_{*}=0.01$, the mass ejection
efficiencies $\xi$ can be large for the same parameters when
$\xi_{met}$ is large, but most of the escaping mass is from the halo,
while the ISM remains bound. (This agrees with MF99 and D'Ercole \&
Brighenti [1999], who found that ISM loss is inefficient so long as it is
distributed non-spherically.)
Thus, a substantial fraction of infalling halo gas can acquire enough
energy from the bubbles to escape the potentials. For example,
halo gas mass comparable to $(\Omega_{b}/\Omega_{0})M_{h}$ escapes from 
the halo potentials with $M_h=5\times10^{8}$ M$_{\odot}$ at $z=8$
and $M_h=5\times10^{9}$ M$_{\odot}$ at $z=3$, even with $f_{*}=0.01$.
These results have two interesting implications for galaxy formation:
further cooling in the halos can be suppressed to a certain degree,
which we discuss in \S\ref{sub:time}; and the
turned-around halo gas will carry the energy of supernova explosions
to other sites of halo formation, perhaps preventing their collapse,
which we discuss next. 

\subsubsection{Energy Transport Efficiencies}
\label{subsub:energy}
In order to compute the energy transport efficiency $\zeta$, we add up
the kinetic energy $K$ and thermal energy $U$ carried by any gas with
$v>v_{esc}$, subtracting the potential energy required to escape the
local potential well:
\begin{equation}
\zeta=\frac{K_{esc} + U_{esc} - (\Phi(2R_v)-\Phi(r))}
{L_m t_f}, 
\label{eej}
\end{equation}
where $\Phi(r)$ is the potential of a gas parcel located at $r$,
$\Phi(2R_v)$ is the potential at $2R_v$ from the center, and the input
energy is $L_m t_f$.  We exclude thermal energy carried by any
swept-up shell of ISM material
and infalling gas, if the Hubble time at a given
redshift $t_H(z)$ exceeds the cooling time of a shell with radius $R_{s}$,
\begin{equation}
\tau_{sh}=(78000\mbox{ yr}) \left(\frac{1+z}{9}\right)^{-3} 
\left[\left(\frac{\nu}{0.2}\right) \left(\frac{f_{*}}{0.01}\right)\right]^{1.7} \left[\frac{R_{s}}{R_v(M_{h},z)}\right]^{0.3} 
\left(\frac{Z}{Z_{\odot}}\right)^{-1}, 
\label{shell50}
\end{equation}
where we assume that the density profile is that of an isothermal halo
[Eq.~(\ref{rho0})], the bubble is driven by the residual energy $\nu
L_m t_f$ with star formation efficiency $f_{*}$, and $R_v$ is the
virial radius for a given halo $M_{h}$ at a given redshift $z$. 
We used 
equation~(\ref{coolingcurve}) for the cooling rate $\Lambda(T)$.
In fact $\tau_{sh} \ll t_H(z)$ unless $f_{*}=0.1$ at $z\le 8$.

We can not systematically distinguish between energy carried by hot,
metal-enriched gas and swept-up, ambient gas because of numerical
diffusion.  However, we can safely compute the amount of thermal
energy carried by escaping shells of swept-up gas with a conservative
color cut-off, $c > 0.6$, because the thermal energy is carried by the
gas which is recently shock heated and therefore not mixed with ejecta
by numerical diffusion.  In Figure~\ref{energyesc}, we plot the
fractions of energy carried by ejecta and by ambient gas with $c >
0.6$ in a halo with $M_h=5\times10^{9}$~M$_{\odot}$ at $z=3$. It shows
that swept-up halo gas carries most of the energy, in kinetic
form. Even if we raise the color cut-off to $c > 1-\epsilon$, the
kinetic energy of the ambient gas remains much larger than that of the
interior gas.  After the bubble blows out at $t\approx10$~Myr, the
thermal energy of the hot interior is effectively transferred to the
surrounding gas.  The bubble reaches $2R_v$ by $t \simeq 100$ Myr, and
the fraction of energy escaping $\zeta \approx 0.18$.  Energy transport
efficiencies $\zeta$ for all the models are listed in
Table~\ref{result}.

We note that numerical diffusion could 
accelerate the cooling of the bubbles. 
However, a resolution study of energy in models with $f_{*}=0.01$ and
$M_h=5\times10^{8}$~M$_{\odot}$ at $z=8$ using standard and
doubled resolution shows no more than 30\% differences in energy
(Fig.~\ref{resolution}). The largest differences are in shell thermal
energy, which is least affected by numerical diffusion.  This suggests
that the dissipation of bubble energy seen in our simulations is
physical, not dominated by numerical diffusion.  

The transfer of the blast-wave energy into the IGM is effective, with
$\zeta\rightarrow1$,
in models with $f_{*}=0.1$ and negligible if
$f_{*}=0.001$, as expected from our discussion of the metal and mass
ejection efficiencies.  The energy transport efficiencies $\zeta$ are
roughly proportional to the mass ejection efficiencies $\xi$, because
the escaping mass is carrying most of the energy.  Both seem to be
largely determined by $f_{*}$.  With $f_{*}=0.01$, the energy transport
efficiencies
are $\zeta\sim0.1$--0.3 in halos
with $v_c\la50$ km s$^{-1}$.
Larger halos with $v_c\approx100$ km s$^{-1}$ 
have more efficient cooling, and likely therefore have higher star formation efficiencies
$f_*\gg0.01$, so that they
can transport more total energy into the IGM.

This kinetic energy feedback could stir intergalactic gas and support
it against gravitational collapse in other halos.  Recent numerical
work has demonstrated that supersonic turbulence can delay
gravitational collapse for many free-fall times in turbulent molecular
clouds (Gammie \& Ostriker 1996; Klessen, Heitsch \& Mac Low 2000).
However, turbulent energy decays
quickly, in a time roughly proportional to the crossing time
$t_{cr}=L/v_{rms}$ of a turbulent region of size $L$ with flow having
{\it rms} velocity $v_{rms}$ (Mac Low 1999; Elmegreen 2000).  The distance
reached by the turbulent motions is only $\sim L \sqrt{2}$ after half
the energy is lost at $t_{cr}$ (Avila-Reese \& V\'azquez-Semadeni
2001). 

{We now crudely estimate whether turbulence driven by starburst
  bubbles in the very first dwarf galaxies could also delay and
  suppress the formation of subsequent dwarf galaxies with the same or
  larger mass
$M_g > M_{J}$. }
{We can estimate how much turbulent energy a single dwarf starburst
galaxy can supply to surrounding overdense regions with baryonic density 
$\rho$, by 
dividing the escaping energy by the volume of the bubble at $R_b$: }
\begin{eqnarray}
v_{rms} & = & \left(\frac{2<\zeta L_m t_f>}{(4/3) \pi \rho R^{3}_b}\right)^{1/2} \\
& \approx & (40\mbox{ km s}^{-1}) 
\left[\left(\frac{\zeta}{0.1}\right)
\left(\frac{f_{*}}{0.01}\right)\right]^{1/2}\left(\frac{\rho/\bar{\rho}}{5}
   \right)^{-1/2}
\left(\frac{R_b}{5R_v}\right)^{-3/2},  
\label{rms}
\end{eqnarray}
where we used equation~(\ref{lmech}) with $t_f=50$ Myr. 
This depends on energy ejection efficiency $\zeta$, star formation
efficiency $f_{*}$, and halo virial radius $R_v=R_v(M_h,z)$ 
from equation~(\ref{rv}). 
Note that the average density of a 
    gravitationally
bound object at turn-around 
$\rho/\bar{\rho}=9\pi^2/16\approx5.6$ from the spherical collapse model. 
{Based on our simulations, 
we estimate $v_{rms}\approx 40\mbox{ km s}^{-1}$ with $f_{*}=0.01$ from 
galaxies with $v_{c}\la 50\mbox{ km s}^{-1}$, 
assuming $\rho/\bar{\rho}=5$ and $R_b=5R_v$.
The same level of turbulence could result from efficient star
formation 
$f_* \sim 0.05$ yielding $\zeta\approx0.02$ in galaxies with 
$v_c\approx100$ km s$^{-1}$.  
Recall that the kinetic energy of the wind is carried by the swept-up halo 
and intergalactic gas and is transformed to turbulence through dynamical
instability. Therefore, 
equation~(\ref{rms}) underestimates the amount of turbulence at the edge of 
the bubble, but overestimates it within the bubble. 

In order to delay and suppress cooling in halos with 
$v_{c}$, the turbulence with $v_{rms}\approx v_{c}$
must be driven and maintained by repeated starbursts occurring 
on a dissipation timescale of the turbulence in equation~(\ref{rms}): 
\begin{eqnarray}
t_{cr}=R_{vir}/v_{rms}\approx 0.3 \mbox{Gyr }
\left(\frac{v_{rms}}{v_{c}}\right)^{-1}
\left[\frac{(1+z)}{4}\right]^{-3/2} \nonumber \\
< t_{H}(z)\approx 1.9\mbox{ Gyr }\left[\frac{(1+z)}{4}\right]^{-3/2}.
\label{tcr}
\end{eqnarray}
Note that some fraction of cooling in the halos with $v_{c}$ 
can be suppressed by $v_{rms}\approx v_{c}/2$, since 
the turbulent pressure delays the turn-around of gas from the Hubble flow.

The
Press-Schechter mass function for our $\Lambda$CDM model 
predicts the mean separation of dwarf galaxies corresponding to $2\sigma$ peaks to be
$\sim10R_{v}$, independent of mass and redshift, when they are assumed to reside in overdense filaments 
with $\rho/\bar{\rho}\approx5$--$10$ 
(corresponding to a volume filling factor of a few percent).
This is likely to be an overestimate, since $2\sigma$ peaks will be clustered. 
In addition, the mean separation 
of all the dwarf galaxies at a given epoch will be smaller, 
so we assume bubbles from these galaxies 
collide with other bubbles at $\sim5R_v$ on average.  
Starburst bubbles are powerful enough to reach 
distances much greater than a few virial radii by $t=50$--100 Myr in 
our models
with $v_c\la 50\mbox{ km s}^{-1}$ and $f_{*}=0.01$ and in all 
models 
with $f_*=0.1$. 
Therefore, we can only expect 
these sorts of halos, which typically form from 1--$2\sigma$ peaks, 
to provide turbulence of the order $\sim40$ km s$^{-1}$ to the IGM. 
Such starburst-driven turbulence may delay and suppress 
the cooling of the gas in 
future halo formation sites. 

The turbulent Jeans mass (Chandrasekhar 1956) is
\begin{equation}
M_{J,t}\approx 2.4\times10^{10} \mbox{M}_{\odot}
\left(\frac{1+z}{4}\right)^{-3/2}  
\left[\left(\frac{c_s}{16~\mbox{km s}^{-1}}\right)^2+\frac{1}{3}
\left(\frac{v_{rms}}{16~\mbox{km s}^{-1}}\right)^2\right]^{3/2},
\label{jeans}
\end{equation}
where $c_s=16\mbox{ km s}^{-1}$ is the sound speed of the IGM with $T=2\times10^{4}$ K.  
{The Jeans mass is 
5.4 times larger 
than the thermal Jeans mass alone for halos
with $v_{rms}=40$~km~s$^{-1}$, corresponding to $v_{c}\approx120\mbox{ km s}^{-1}$. 

This mechanism is analogous to the suppression of dwarf galaxies by UV 
background radiation 
(Bond, Silk \& Szalay 1988; Meiksin 1994; 
Vedel et al.\ 1994; Thoul \& Weinberg 1996; Navarro \& Steinmetz 1997; 
Iliev, Raga \& Shapiro 2003),
except that the gas 
is energized by turbulent flows rather than thermal pressure due to ionizing radiation. 
Semi-analytical studies of galaxy formation suggest that 
suppressing the fraction of gas that can cool and form stars
below 
$\sim10$--20\% in halos with $v_{c}\approx50$--100~km~s$^{-1}$ 
can reproduce
the shape of the observed galaxy luminosity function (e.g. Cole et al.\ 1994).}


Some cosmological simulations have demonstrated that the galactic
outflows that they model, though not based on superbubble dynamics,
can suppress the formation of dwarf galaxies by heating up the gas by
shocks (Theuns et al.\ 2002) and by stripping the gas from the
potentials by their momentum (``baryonic stripping'' Scannapieco et
al.\ 2001).  {We agree that starburst-driven turbulent flows may
play an important role in regulating the formation of galaxies, and 
so may be a key to the long-standing problems of 
overcooling and angular momentum losses in a CDM universe. 
It appears useful to model such starburst-driven turbulence 
in more detail in cosmological simulations, in order to address such 
questions.} 


In terms of the locally induced metallicity $Z$, the
{\it rms} turbulent velocity is
\begin{equation}
v_{rms}\approx (60\mbox{ km s}^{-1})
\left(\frac{\zeta}{\xi_{met}}\right)^{1/2}
\left(\frac{Z}{10^{-2}\mbox{ Z}_{\odot}}\right)^{1/2}.
\label{rmsZ}
\end{equation}
The trend toward larger galaxy masses dominating metal ejection at later times 
(see Figure~\ref{Zmetal}) suggests that $v_{rms}$ may increase with decreasing redshift. 
The values found for $v_{rms}$ are comparable to the level of kinematic disturbance measured
in C{\sc iv} systems (Rauch et al. 2001a), suggesting that the metal deposition
may have been recent in these systems, before the turbulence had
time to dissipate.
For a lower mean intergalactic metallicity of
$Z=0.001 \mbox{ Z}_{\odot}$, $v_{rms}\approx20$ km s$^{-1}$, which
could result in significant broadening of the Ly$\alpha$ absorption
systems and help account for the discrepancy between the predicted
and measured line widths detected even in the optically thin systems
(Meiksin, Bryan \& Machacek 2001). On the other hand, there would likely
have been adequate time for the 
turbulence
to dissipate by the time
the winds could reach the underdense regions where the optically thin systems
are expected to reside. Although the energy of the turbulence may be
converted into heat as the turbulence dissipates, it is unclear how efficient
this would be. Over a local Hubble time 
$t_H(z)$, 
the energy associated with the
turbulence, even if converted to heat, would also decay due to cosmological
expansion. This is a topic that requires further exploration.

\subsection{Pressure Confinement of Bubbles}
\label{sub:press}
All of the bubbles in our models are confined by external pressures if
$f_{*}=0.001$, while all escape to the IGM if $f_{*}=0.1$.  These
results agree with our prediction in Figure~\ref{rpcon} if the
fraction of energy preserved in the bubbles $\nu \la 0.2$.  For
detailed analysis of how ram pressure affects the evolution of the
bubbles, we choose models with $M_{h}=5\times10^{9}$~M$_{\odot}$
and $f_{*}=0.01$ at $z=3$, 5, and 8, in which the 
efficiency of metal and mass ejection and of energy transport 
decrease steadily as a function of redshift.
We also try to constrain $\nu$ in equation~(\ref{rpm}) by comparing
our results with Figure~\ref{rpcon}.

We plot in Figure~\ref{rpg95} the initial total pressure distributions as a
function of radius in these three halos. We examine the
sum of thermal and ram pressure because some kinetic energy is
converted to thermal energy in shocks because of the artificial
turn-off of cooling in the central regions,
%
%
but the total energy is still conserved since we do not solve for
collisional cooling in the infall code (see \S\ref{sub:halo}). 
%
We also plot the total pressure distributions of bubbles at 45 degrees
from the plane of the disks, measured at $t=85$ Myr, for the halos at $z=3$ 
and 8. The ram pressure is included only for the non-colored gas 
moving with positive radial velocities. 

Figure~\ref{rpg95} shows that the 
total pressure in the undisturbed halos in our models
is a factor of 
five lower at $r\la R_v$ than the ram pressure 
of the infalling halo gas
calculated from equations~(\ref{vin}) and (\ref{rho0}) in \S\ref{sec:predict},
but the difference is less than a factor of two outside $R_v$.  Since
pressure confinement mostly acts at $r\ga R_v$ for powerful bubbles
with $f_{*}\ga0.01$, we can reasonably compare our results with the
prediction made in \S\ref{sec:predict}.  The external pressure
increases as a function of redshift, consistent with the prediction
from equation~(\ref{vin}): $\rho v^2_{in}\propto (1+z)^4$.

Now we try to constrain the bubble energy fraction $\nu$.
All the bubbles in models with $f_{*}=0.01$ expand beyond $R_v$ by
$t=50$ Myr when the starbursts end, except along the planes of the
galaxies where the ISM offers the most resistance
(Fig.~\ref{figures}{\em b}).  The bubbles generally grow to $r=2R_v$
by $t\approx 100$ Myr, leaving most of the ISM and the fragmented
shells of swept-up halo gas behind.  The halo pressure has already
begun to confine the bubble at $z=8$ by $t=85$ Myr (Fig.~\ref{rpg95}).
Only a small fraction of original bubble energy remains at the swept-up shells
in the form of kinetic energy. As a result, metal, mass, and energy feedback efficiencies
are less than a few 
percent.  
Pressure confinement begins 
much later at $z=5$,
at $t\approx150$ Myr, after 
the bubble 
grows beyond $2R_v$.  Since we assume that
the low-density IGM begins at $2R_v$, we compute moderately high
ejection efficiencies of metals and halo mass of
$\sim0.3$--0.4 in this
case.  The pressure on the bubble at $z=3$ is more than an order of
magnitude less than
at $z=8$ (Fig.\ \ref{rpg95}), so it expands
beyond $2R_v$ without any significant influence from the infall,
expelling most of the metals and returning a large fraction of infalling gas
into the IGM.

These halos fall just around the line of pressure confinement with
$\nu f_* =1\times10^{-3}$ in Figure~\ref{rpcon}.  We compute $\nu$ in our models by summing
the energy of interior gas with $c<0.6$ and swept-up gas with
positive velocity, and dividing by $L_m t_f$.  More than half of the
injected energy has already been lost in these halos by $t=50$
Myr. Loss occurs by radiative cooling in the swept-up shells and in
the mixed gas after the shells fragment. It continues as time
progresses and is larger at higher redshift. Our resolution study
shows that the dissipation of bubble energy seen in our simulations is
physical, not dominated by numerical diffusion (see Fig.\
\ref{resolution}).  

We find the average energy fraction
to be $\nu \sim0.2$ between $t=50$ and 100 Myr in the halos 
with $M_h=5\times10^{9}$ M$_{\odot}$ and $f_*=0.01$ at $z=3$ and 5
and to be $\nu < 0.2$ at $t=50$ Myr and $\nu \la 0.05$ by
$t=85$ Myr in the same halo at $z=8$. 
Therefore, we have $\nu\approx0.1$--0.2 from our simulations for
$f_{*}=0.01$, and so our results are consistent with the prediction
made in Figure~\ref{rpcon}.  We can also use equation~(\ref{tconf}) to
define the confinement criterion in terms of virial temperature
or circular velocity instead, finding that halos with $T_v \ga 1.0 \times 10^5$~K 
or $v_c\ga52$ km s$^{-1}$ will begin to confine
a starburst with $f_* = 0.01$ and $\nu\approx0.1$. 
On the other hand, $\nu$ increases more
than a factor of two to $\nu\ga0.6$ in the same halos when
$f_{*}=0.1$.  This is because the cooling time of the swept-up shells
$\tau_{sh}$ becomes larger than the dynamical time of the bubbles
within $2R_v$, so the shocks remain adiabatic, and the fragmentation
of the shells is suppressed.  Bubbles even in galaxies with $v_c\sim280$ km s$^{-1}$ 
cannot be stopped by ram pressure if $f_{*}=0.1$ with $\nu\ga 0.6$.

In summary, the ram pressure of the infalling halo gas strongly
influences feedback efficiency in dwarf galaxies, although it is
primarily determined by the strength of star formation.  The analytic
results of \S\ref{sec:predict} appear to offer a reliable guide to the
importance of ram pressure.

\subsection{Fallback Timescales}
\label{sub:time}

We compute the time for swept-up ISM and halo gas to cool and fall
back to the center, which gives a minimum time before the next
starburst.  We define the accretion fraction as
\begin{equation}
\Upsilon(t) = M_4(R, t) / M_4(R_{*}, 0), 
\label{acc}
\end{equation}
where $M_4(R,t)$ is the gas mass with $T<10^{4}$~K within radius $R$
at time $t$, and $R_*$ is the size of the galactic disk. 
Since we run our simulations for 200 Myr after the onset of
starbursts or until the bubbles leave the grids, we first compute
$\Upsilon(t)$ up to the end of the runs with $R=R_{*}$.

If most of the gas is bound to the potential at that time, but hasn't yet
fallen back to the center,
we calculate $\Upsilon(t)$ for later
times using a 
one-dimensional,
ballistic approximation with $R=0$. Bound gas parcels with
$v<v_{esc}$ are assumed to travel on radial, ballistic orbits in the
potential, ignoring non-radial motions and acceleration of the accretion
by cooling. We apply the approximation to gas within $2R_v$ at the
time that feedback efficiencies are computed. This simple ballistic
approximation was used by Zahnle \& Mac Low (1995) to follow the
ejecta of a typical Shoemaker-Levy 9 impact falling back onto
Jupiter's atmosphere, and was found to give results consistent with
observations.  We also tested the validity of the method by computing
the accretion fraction $\Upsilon(t)$ based on the ballistic
approximation in a model in which we also compute
$\Upsilon_{sim}(t)$ directly in the simulations. The bottom panel
of Figure~\ref{acm} shows that they
agree within $\sim 5\%$.  
We find similarly good agreement for other cases in which
$\Upsilon_{sim}(t)$ approaches unity so that we can make the comparison.

Figure~\ref{acm} also shows the accretion fractions $\Upsilon(t)$ as a
function of time for three halos with $M=5\times10^{9}$ M$_{\odot}$ at
$z=3$, 5, and 8, with $f_{*}=0.01$ and 0.001.
Because the ballistic model is only a one-dimensional approximation,
it does not give exactly the same answer as the numerics at low values
of $\Upsilon$, as seen in Figure~\ref{acm}.
We also define the
fallback time $\tau_f$ as the time when $\Upsilon(\tau_f)=e^{-1}$, and
give $\tau_f$ for all our model halos in Table~\ref{tacc}.
%
The bump in $\Upsilon(t)$ for the $z=8$ model occurs when ISM material
falls back to the center almost at the free-fall velocity, creating accretion
shocks.  

Since none of the mass escapes in models with $f_{*}=0.001$,
$\Upsilon(t)$ quickly approaches unity within $0.2t_H(z)$.
With $f_{*}=0.01$, feedback is more efficient in halos at lower
redshift, but the timescale for gas to fall back remains $\la
0.4t_H(z)$.  This is rather significant in halos at $z=3$ and 8 
that
have measured mass ejection efficiencies $\xi\approx1$.  Most disk gas appears
tightly bound in galaxies with star formation efficiencies $f_{*}\la
0.01$.  In such galaxies, subsequent starbursts will occur in
pre-enriched gas with a fraction $(1-\xi_{met})$ of the metals from
previous starbursts, yielding 
\begin{equation}  
Z\approx0.06\left(\frac{1-\xi_{met}}{0.5}\right)\left(\frac{f_*}{0.01}\right)Z_{\odot}.
\end{equation}
The presence of processed metals probably
enhances star formation efficiency, which is the dominant factor
influencing stellar feedback in our simulations.  If the initial
starbursts are very large, with $f_{*}=0.1$, the mass in the halos is
blown away, and does not fall back within the local Hubble time
$t_H(z)$.  Of
course, real galaxies may never reach this star formation efficiency,
as we discussed in \S\ref{sec:disk}.

Our results suggest that it is difficult to keep most of the gas in a
dwarf halo hot and diffuse and to prevent the central concentration of
gas, once a dense disk-like object has formed. Star formation in small
dwarf galaxies may occur in the form of repeating starbursts, although
some other mechanism must explain the pauses longer than $10^9$ years
between observed bursts of star formation.  

If feedback is to solve the overcooling and angular momentum problems
in the formation of massive galaxies, it must prevent the initial
collapse and concentration of gas into dwarf galaxies in the halos of
the massive galaxies. Once much of the gas in the massive halo has
cooled to form dwarf galaxies, subsequent starbursts in the dwarf
galaxies have great difficulty heating and expelling it permanently.
The dwarf galaxies then lose angular momentum from dynamical friction
against the parent dark halo. Feedback from dwarf galaxies may prevent gas in
nearby dwarf halos from collapsing, however, leaving a reservoir of
diffuse gas to build a massive galaxy without the baryons 
losing too much angular momentum through dynamical friction.
 
\section{Conclusion}
\label{sec:concl}

We study feedback from dwarf starburst galaxies with 
virial temperatures
$T_{v}>10^{4}$~K at high redshift, by numerically modeling the interaction of
superbubbles driven by repeated supernova explosions with the ram
pressure from cosmological infall. We compute ejection efficiencies for
mass $\xi$, metals $\xi_{met}$, and energy
$\zeta$, after the bubbles have expanded well beyond the virial radius,
and estimate the timescale for disturbed gas to fall back and
become available for subsequent starbursts.
We have tried to consistently choose approximations that underestimate
the effects of supernova feedback, so that our computations of the
efficiency of mass, metal, and energy feedback and fallback time
should be reasonably strong lower limits to the actual values.
We now list our conclusions.

\begin{itemize}

\item~Ejection and transport efficiencies are primarily determined by the efficiency
of star formation $f_{*}$.  With $f_{*}=0.1$, 
nearly all the metals produced in the starburst escapes
($\xi_{met}\approx1$), 
most of the disk ISM
and large parts of the infalling halo are blown away ($\xi>1$),
and accelerated halo gas carries the energy
of the bursts out to the IGM ($\zeta>0.5$). With $f_*=0.01$, moderate to
high feedback efficiencies are only observed in halos with $v_c\approx30$--50 
km s$^{-1}$.
With $f_{*}=0.001$, none of the metals, the mass, nor
the energy can escape the potentials.

\item~The ram pressure of infalling halo gas can suppress the growth
of starburst bubbles and thus prevent the metal, mass, and
energy feedback, unless  $f_{*}
\ga 
0.1$.  
The energy available to drive the bubble depends on both $f_*$ and on the fraction of energy 
in the hot gas actually available to drive the bubble $\nu$.
The minimum halo mass for ram pressure
confinement with a given $\nu f_{*}$ can be specified by a 
cut-off in 
virial temperature $T^{*}_{v}$ or circular velocity $v^{*}_c$ 
at any redshift (Eq.~\ref{tconf}).
Our prediction is consistent with the values of $\nu$, $\xi_{met}$, $\xi$, 
and $\zeta$ found in our simulations for a given $f_*$. 
With $\nu f_{*}=1\times10^{-3}$, we can only expect high feedback efficiencies in
halos with $T_{v}\la T^{*}_{v}\approx 1\times10^{5}$ K and $v_c\la v^{*}_c\approx
52$ km s$^{-1}$, above the limit for hydrogen line cooling, 
and below the Jeans mass in the reheated IGM.

\item~We find that small dwarf galaxies with $v_c\approx30$--50 km s$^{-1}$
at $z\la8$ are efficient in enriching the IGM with metals, yielding 
$\bar{Z}\ga10^{-3}$ Z$_{\odot}$ with $f_*=0.01$. 
The early metal enrichment may be
able to explain the absence of kinematic disturbance observed in
low-density Lyman $\alpha$ clouds.
Larger dwarf galaxies with
$v_c\approx100$ km s$^{-1}$ at $z=3$--5
might
enrich the IGM more effectively, 
yielding $\bar{Z}\approx10^{-2}$ Z$_{\odot}$, because efficient cooling 
expected in such halos 
may
enable $f_{*}\gg0.01$. The late metal enrichment 
is consistent with high metallicities and large turbulent motions 
observed in high-density Lyman $\alpha$ clouds. 
These galaxies form out of 1--2$\sigma$ peaks in our $\Lambda$CDM Gaussian density 
perturbation. 

\item~We expect mass ejection efficiencies $\xi\approx0.3$--1 and energy transport
efficiencies $\zeta\approx0.1$-0.3 from the dwarf galaxies 
with $v_c\approx30$--50 km s$^{-1}$ with $f_*=0.01$. However, most of the ISM remains bound
unless $f_{*}=0.1$. Instead, a large amount of infalling gas is
turned around by the galactic wind and escapes the potential. This
outflow may carry kinetic energy
%
into the IGM, broadening Ly$\alpha$ absorption systems, and
%
to other nearby halos to prevent gravitational collapse of the gas in them. 
We estimate
that the ensemble of outflows may provide turbulent support with
typical $v_{rms}\approx20$--40 km s$^{-1}$.  Further study of
this multiscale problem is required to determine whether this
mechanism of suppressing dwarf galaxy formation can solve the
overcooling and angular momentum problems.

\item~The timescale for swept-up mass to fall back to the center
of a halo is short
compared to the Hubble time, $\tau_f \la 0.4 t_H(z)$
if $f_{*}\la0.01$. We expect star formation
efficiencies for subsequent starbursts to be larger than those of the
initial starbursts that we studied, because ejected metals bound
in the potentials will enhance gas cooling. 
%
Once gas in a halo cools to form a disk-like object, it is difficult
to blow away the disk gas. 

\end{itemize}

We must mention the caveats in our study:

Our models assume an instantaneous starburst at the center of a
galactic disk.  This is obviously an oversimplification, and such
instantaneous starbursts, especially with $f_{*}\gg0.01$ are probably
not realistic.  We instead expect multiple OB associations or star
clusters to form over a few hundred pc across the disk (e.g. Vacca
1996; Martin 1998).  Each cluster will likely behave like our models
however, so the net effect will be greater than we compute, consistent
with our attempt to present lower limits to the effects of stellar
feedback. 
Given the dominant influence of the star formation efficiency on the
importance of stellar feedback, a self-consistent, physically
motivated modeling of star formation will be required to further our
understanding.

We entirely neglected magnetic fields in this study. Magnetic
fields can inhibit the formation of cold, dense shells, and suppress
the fragmentation of the shells, perhaps reducing or preventing the
dissipation of the bubble energy through the fragmented shells.  Then
the fraction of energy preserved in the bubbles $\nu$ might be larger,
giving higher feedback efficiencies in the dwarf galaxies that we
modeled.  On the other hand, magnetic pressure and tension might act
to help confine the expanding bubbles (e.g.\ Tomisaka 1998), depending
on the strength of the field.

Our models are not three-dimensional.  This has consequences for two
main reasons.  First, we compute the Rayleigh-Taylor instability in
fragmenting shells during blowout in two dimensions. This has the
effect of increasing the size of the fragments (Mac Low et al.\ 1989)
because only ring-like modes of the instability can form.  In three
dimensions, the instability takes on a spike and bubble form, with the
spikes representing the dense fragments.  This does not change the
gross topology, since even in two dimensions the shell still overturns
and the wind can escape.  However, a detailed study of the fragment
properties cannot be done.  As we restrict ourselves to integrated
quantities, however, this probably does not strongly affect our
results. Certainly the two models computed by Wada \& Venkatesan (2003)
in three dimensions appear to grossly agree with our results: a star
formation efficiency of $f_* = 0.14$ in a galaxy with
$10^7$~M$_{\odot}$ of gas leads to metal escape, while with $f_* = 0.014$
the metals remain confined.

Second, our models assume spherically symmetric gas infall, and the
presence of a well-formed, rotationally supported disk. 
 It is also an oversimplification to compute metal, mass, and energy 
feedback efficiencies when the bubbles reach $R_s=2R_v$.
Realistically, a halo is embedded in the complicated
weblike structures predicted by three-dimensional hydrodynamic
simulations (e.g.\ Miralda-Escude et al.\ 1996; Zhang et al.\ 1998).
Even inside its virial radius, gas accretion occurs along filaments,
and the accretion shock remains highly aspherical even after the
protogalaxy has collapsed to less than a few times the virial radius, 
(Abel et al.\ 1998; 2000; 2002). 
In addition, a rotationally supported disk may
not form before a significant amount of star formation occurs.  We
note, however, that a more dynamical, filamentary background will
create funnels through which superbubbles and galactic winds can
freely expand, potentially increasing the efficiency of metal ejection and
energy transport. 


Although these caveats identify important points requiring further
investigation, none of them appear to invalidate our results; nor do
they suggest that our lower limits are drastic underestimates of the
actual effects of stellar feedback.

\acknowledgements
We thank T. Abel, M. Norman,
J. Sommer-Larsen, V. Springel, and E. Tolstoy for useful discussions, and 
S. Glover, S. Kahn, and Z. Haiman for careful 
reading
of the manuscript. We also thank B. Ciardi for providing useful data to us.  
AF was supported by the Beller Fellowship of the American Museum of
Natural History. She thanks the Theoretical Astrophysics Center in
Copenhagen for hospitality during work on this project.  M-MML was
partly supported by NSF 
grants AST 99-85392 and AST 03-07854.
A. Ferrara acknowledges support from  the RTN
Network `The Physics of the Intergalactic Medium'
set up by the European Community
under the contract HPRN-CT2000-00126 RG29185.
A. Meiksin thanks the University of Edinburgh Development Trust for support.
Computations were
performed on the SGI Origin 2000 machines of the Hayden Planetarium and of
the National Center for Supercomputing Applications.  We thank
M. Norman and the Laboratory for Computational Astrophysics for use of
ZEUS.  This research made use of the Abstract Service of the NASA
Astrophysics Data System.

\newpage

\clearpage

\begin{table}   
\begin{center}
\caption{Model galaxy properties
\label{scale} }
\begin{tabular}{cccccccccc}
   $z$\tablenotemark{a}      & $M_{h}$\tablenotemark{b}  &
   res\tablenotemark{c}      & $R_v$\tablenotemark{d} &
   $R_{*}$\tablenotemark{e}   & $T_{4,v}$\tablenotemark{f}        &
   $H$\tablenotemark{g}       & $v_{esc}$\tablenotemark{h}  & $v_c$\tablenotemark{i} \\ 
   & (M$_{\odot}$)      & (pc)   & (kpc)     & (kpc)   & ($10^{4}$ K) & (pc) & km s$^{-1}$ 
& km s$^{-1}$ \\
\hline
13 & $5\times10^8$  & 2.7        & 1.6       & 0.17    &  4.7        & 35    & 48 & 37 \\    
8  & $5\times10^8$  & 5.2        & 2.5       & 0.27    &  2.0        & 74    & 40 & 29 \\
8  & $5\times10^9$  & 5.6        & 5.5       & 0.58    &  14.        & 220   & 74 & 62 \\   
5  & $5\times10^9$  & 6.9        & 8.2       & 0.87    &  9.4        & 170   & 65 & 50 \\
5  & $5\times10^{10}$ & 15.      & 18.       & 1.9     &  43.        & 360   & 130 & 110 \\
3  & $5\times10^9$  & 10.        & 12.       & 1.3     &  6.2        & 400   & 52  & 41 \\
3  & $5\times10^{10}$ & 25.      & 26.       & 2.8     &  29.        & 820   & 106 & 88 \\
\end{tabular}
\tablenotetext{a}{Redshift}
\tablenotetext{b}{Halo mass}
\tablenotetext{c}{Central zone size}
\tablenotetext{d}{Virial radius}
\tablenotetext{e}{Size of source region}
\tablenotetext{f}{Virial temperature}
\tablenotetext{g}{Scale height of galactic disk}
\tablenotetext{h}{Escape velocity}
\tablenotetext{i}{
    Circular
velocity}
\end{center}
\end{table}

\begin{table} 
\begin{center}
\caption{Metal and mass ejection, and energy transport efficiencies
\label{result} }
\begin{tabular}{cllcccc}
$z$ & $M_{h}$ & $f_{*}$\tablenotemark{a}  & $\xi_{met}$\tablenotemark{b} & $\xi$\tablenotemark{c} & $\zeta$\tablenotemark{d} & Blow-Away\tablenotemark{e}  \\
  & (M$_{\odot}$) &   &   & &   & \\
\hline

13 & $5\times10^{8}$   & 0.1    & 0.99 & 3.2     & 0.45    & Yes            \\
   &                   & 0.01   & 0.68 & 0.33      & 0.17    & No       \\
   &                   & 0.001  & 0.0  & 0.0       & 0.0     & No       \\

8  & $5\times10^{8}$   & 0.1    & 0.99 & 4.5     & 0.76    & Yes       \\
   &                   & 0.01   & 0.64 & 1.1     & 0.34    & No       \\
   &                   & 0.001  & 0.15 & 0.026     & $5.5\times10^{-4}$ & No       \\
   & $5\times10^{9}$   & 0.1    & 1.0  & 1.7     & 0.48    & Yes     \\
   &                   & 0.01   & 0.03 & 0.0088    & 0.024   & No     \\
   &                   & 0.001  & 0.0  & 0       & 0.0     & No     \\

5  & $5\times10^{9}$   & 0.1    & 1.0  & 1.8     & 0.80    & Yes    \\
   &                   & 0.01   & 0.44 & 0.30      & 0.069   & No   \\
   &                   & 0.001  & 0.0  & 0.0       & 0.0     & No \\

3  & $5\times10^{9}$   & 0.1    & 1.0  & 1.7     & 0.92    & Yes         \\  
   &                   & 0.01   & 0.88 & 0.99      & 0.18    & Yes         \\
   &                   & 0.001  & 0.0  & 0.0       & 0.0     & No         \\  
5  & $5\times10^{10}$  & 0.1    & 1.0  & 0.86      & 0.73    & No           \\
   &                   & 0.01   & 0.0  & 0.0       & 0.0     & No            \\
   &                   & 0.001  & 0.0  & 0.0       & 0.0     & No            \\
3  & $5\times10^{10}$  & 0.1    & 1.0  & 1.2     & 0.75    & Yes     \\
   &                   & 0.01   & 0.0  & 0.0       & 0.0     & No     \\
   &                   & 0.001  & 0.0  & 0.0       & 0.0     & No \\
\end{tabular}
\tablenotetext{a}{Star formation efficiency}
\tablenotetext{b}{Metal ejection efficiency}
\tablenotetext{c}{Mass ejection efficiency}
\tablenotetext{d}{Energy transport efficiency}
\tablenotetext{e}{
Whether more than half of ISM has been blown away from galactic potential
}
\end{center}
\end{table}

\begin{table}
\begin{center}
\caption{Estimated fallback time $\tau_{f}$
\label{tacc}
}
\begin{tabular}{cccccc}
$z$     & $M_{h}$     & 
                        \multicolumn{3}{c}{$\tau_f$\tablenotemark{a}} & 
                $t_H$\tablenotemark{b} \\
        & (M$_{\odot}$)   & 
                             \multicolumn{3}{c}{(Myr)}
                                                       & (Myr)   \\
$f_{*}$ &             &  0.001      & 0.01          & 0.1 & \\
\hline
13 & $5\times10^8$    & 65    & 110    & 230   & 290       \\
8  & $5\times10^9$    & 91    & 140    & $\gg t_H$ & 570  \\   
   & $5\times10^8$    & 127   & 160    & $\gg t_H$                &     \\
5  & $5\times10^{10}$ & 116   & 280    & $\gg t_H$                 & 1000 \\
   & $5\times10^9$    & 109   & 320    & $\gg t_H$                 & \\
3  & $5\times10^{10}$ & 320   & 460    & $\gg t_H$                 & 1900 \\
   & $5\times10^9$    & 360   & 740    & $\gg t_H$ & \\
\end{tabular}
\tablenotetext{a}{Estimated fallback time 
such that
$\Upsilon(\tau_{f})=e^{-1}$}
\tablenotetext{b}{Hubble time
at given redshift}
\end{center}
\end{table}

\clearpage

\begin{figure}
\plotone{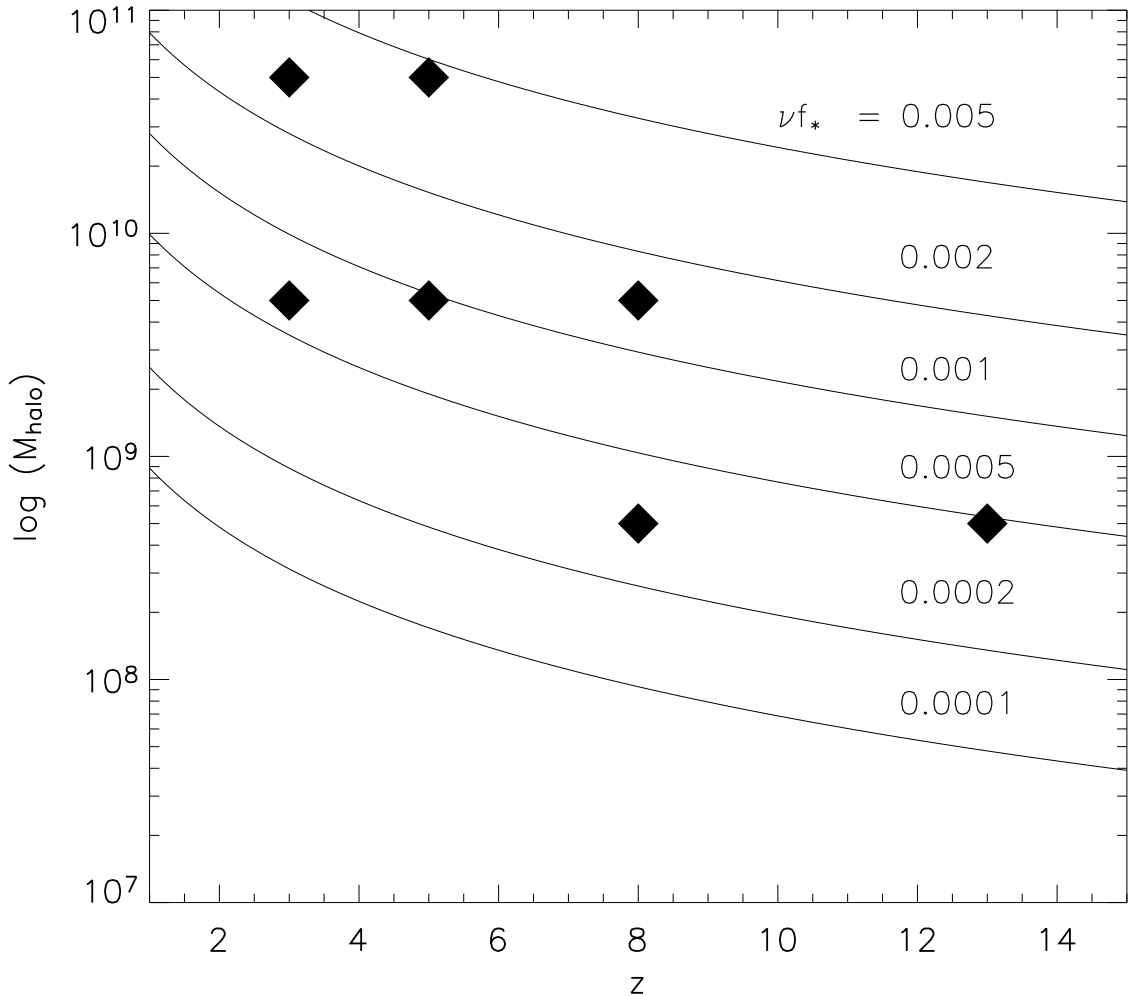} \figcaption{We plot the minimum halo mass for
ram-pressure confinement of bubbles as a function of redshift for 
$0.0001\le \nu f_*\le 0.005$, where a
fraction of injected energy preserved in the bubbles is $\nu$ and 
star formation efficiency is $f_*$.  The galaxies we choose to model
(described below in section~\ref{sub:galaxies}) are shown in filled
diamonds. The starburst bubbles in all
our model galaxies are predicted to escape to intergalactic space if
$\nu f_{*}\ga0.005 $ and to be confined by ram pressure of the infalling
gas if $\nu f_{*}\la 0.0005$.
\label{rpcon}
}
\end{figure}

\begin{figure}
\plotone{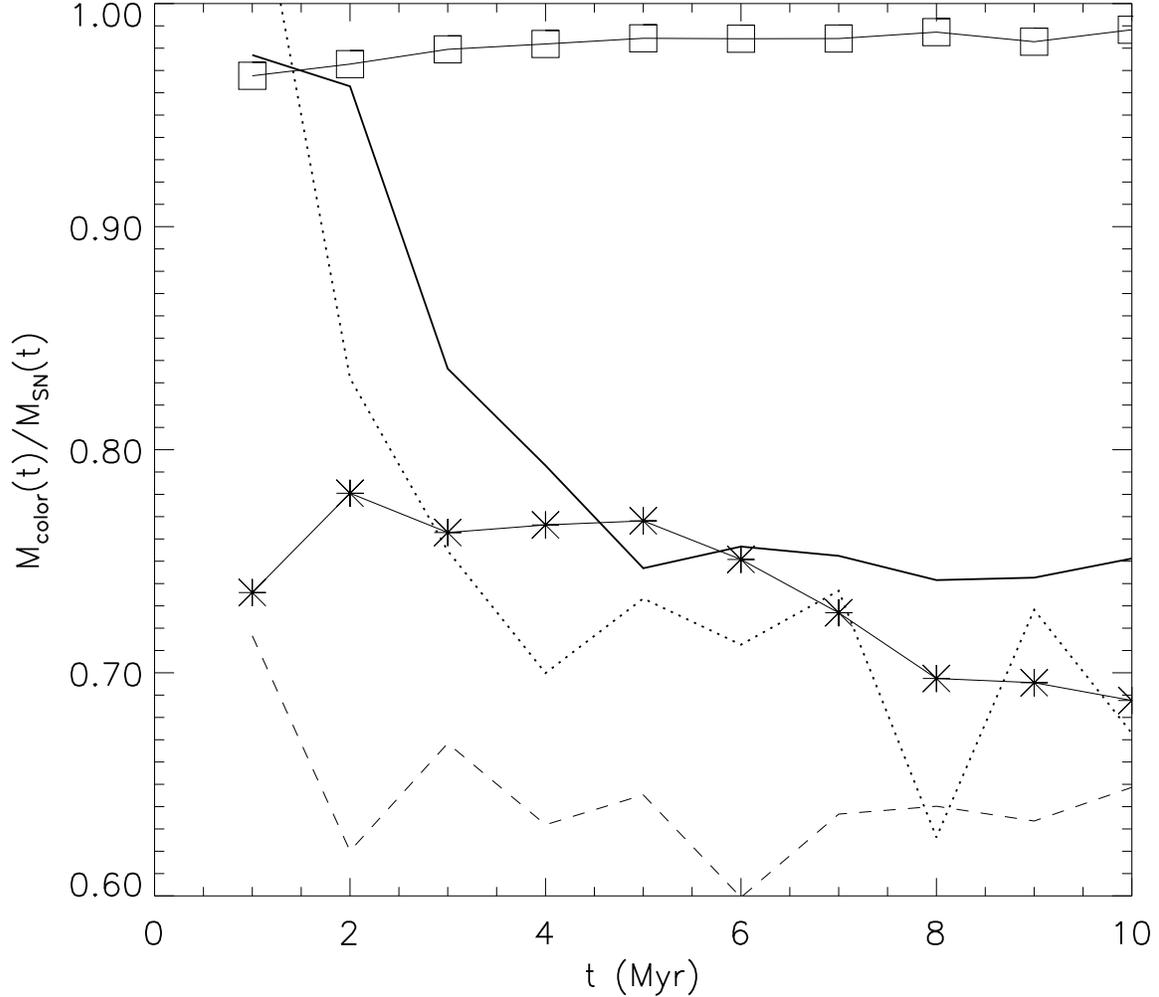} \figcaption{ We show the fraction 
of mass tracked by the tracer field with $c<0.6$, $M_{color}(t)$, over the
total mass injected with $c=0$, $M_{SN}(t)$, in our test simulations as a function
of time up to 10 Myr.  The results are based on 
a standard model at $z=13$, 
with halo background density $\rho_{bg} = 200 \rho_0$, $L_m=10^{40}$ erg s$^{-1}$, the
primordial radiative cooling $\Lambda_{line}$ and a resolution of 4 pc
({\em thick solid line}), the C100 model with extra cooling,
$100\Lambda_{line}$ ({\em solid line with stars}), the C0 model with no
cooling ({\em solid line with squares}), the D10 model with the background
density 10 times higher than the standard model ({\em dashed line}), and the
LR model with zones four times larger than the standard model
({\em dotted line}).  The tracer field fails to pick up some of the injected
gas due to the unresolved, high-density shells at the contact
discontinuity.
\label{ftracer}
}
\end{figure}

\begin{figure}
\plotone{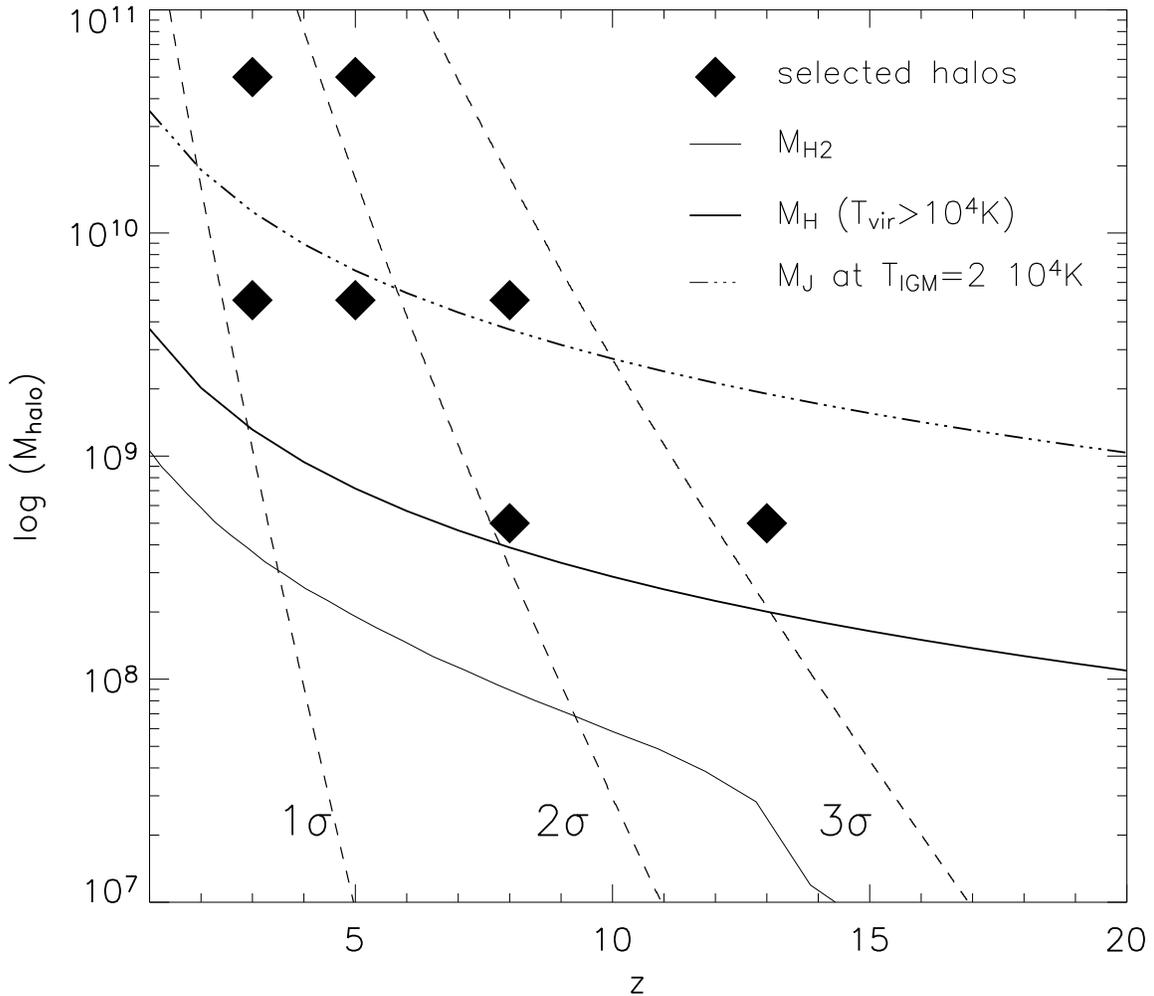}
\figcaption{We choose seven second generation systems 
({\em filled diamonds}) with virial temperature $T_{v}>10^{4}$ K ({\em thick solid line}).
As a comparison,
the minimum halo mass in which
the gas can cool through molecular hydrogen is shown in {\em solid line},
 computed by Tegmark et al.\ (1997) and given by B. Ciardi. 
The Jeans mass after reionization is plotted in {\em dash-dot-dot line}. 
We also plot halo mass expected to collapse from $\Lambda$CDM density
perturbations as $1\sigma$, $2\sigma$, and $3\sigma$ peaks ({\em dashed line}).
\label{selh}
}
\end{figure}




\begin{figure}
\epsscale{1.0}
\plotone{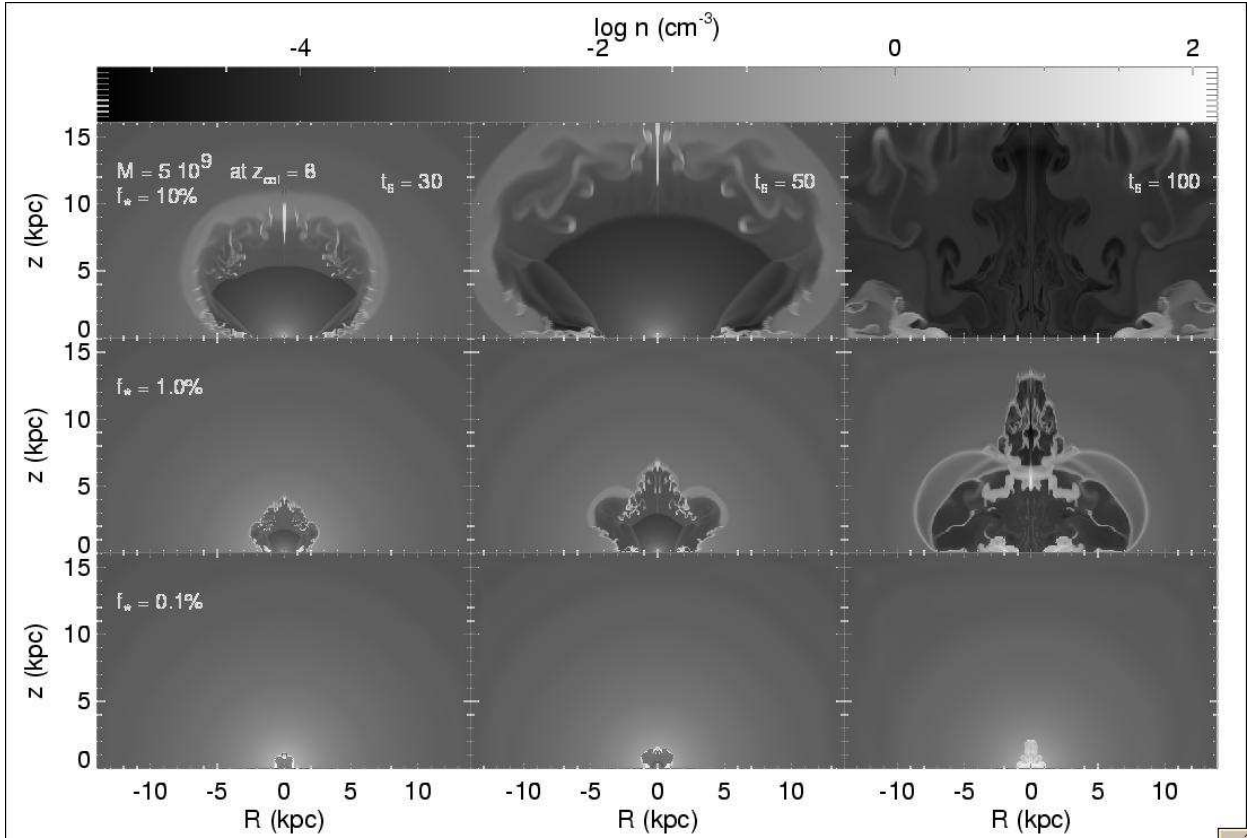}
\caption{Density distributions for our model with halo mass
$M_h=5\times10^{9}$~M$_{\odot}$ at $z=8$ are shown at times 30 Myr
(left), 50 Myr (middle), and 100 Myr (right), after the onsets of
starbursts with star formation efficiencies $f_{*}=0.1\%$ (bottom),
1\% (middle) and 10\% (top). Corresponding mechanical luminosites are
$L_m=1.8\times10^{39}$, $1.8\times 10^{40}$, and $1.8\times
10^{41}$~erg~s$^{-1}$ respectively.
\label{z8g95}
}
\end{figure}

\begin{figure}
\epsscale{1.0}
\plotone{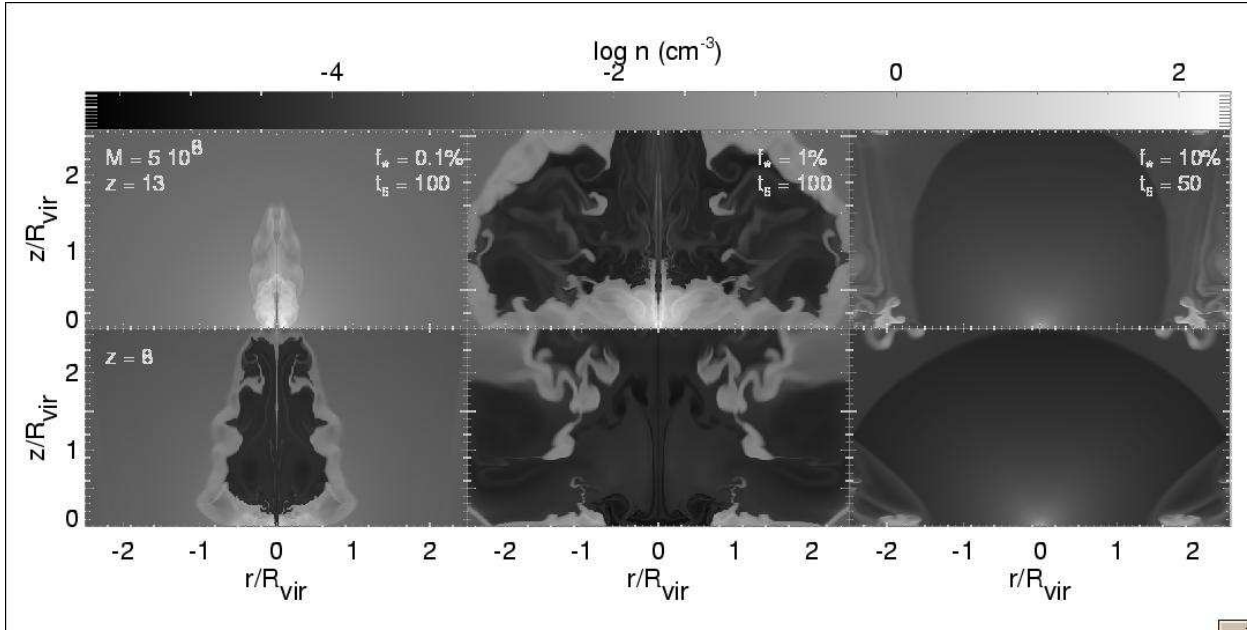}
\caption{Density distributions at $t=100$ Myr for $f_{*}=0.1\%$ (left
panels) and 1\% (middle panels) and at $t=50$ Myr for $f_{*}=10\%$
(right panel) for models with halo mass 
({\em a}) $M_h=5\times 10^{8}$~M$_{\odot}$ at $z=13$ and 8, 
({\em b}) $M_h=5\times 10^{9}$~M$_{\odot}$ at $z=8$, 5, and 3, and 
({\em c}) $M_h=5\times10^{10}$~M$_{\odot}$ at $z=5$ and 3.
\label{figures}
}
\end{figure}

\addtocounter{figure}{-1}
\begin{figure}
\epsscale{1.0}
\plotone{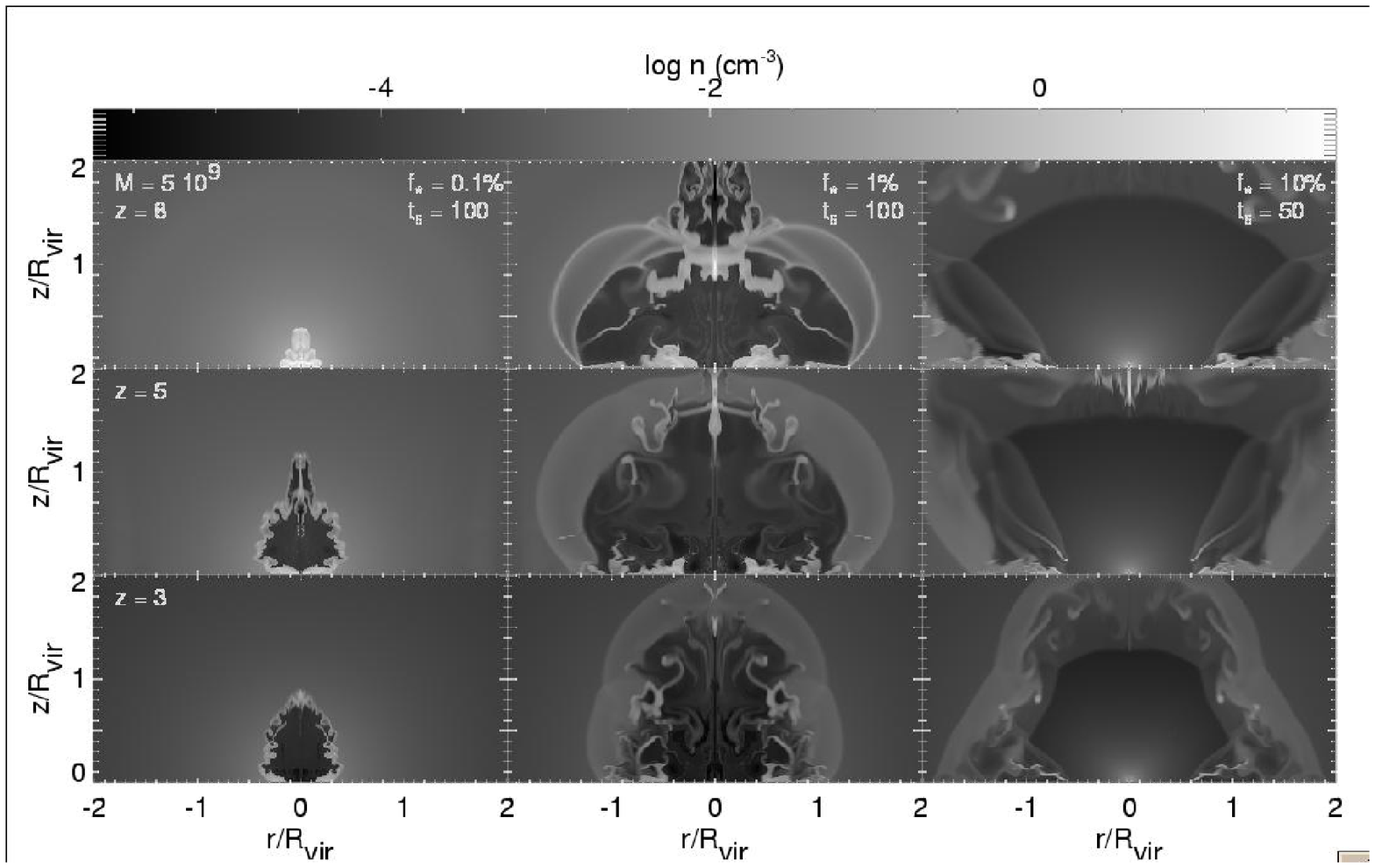}
\caption{
Figure~\ref{figures}{\em b)}
}
\end{figure}

\addtocounter{figure}{-1}
\begin{figure}
\epsscale{1.0}
\plotone{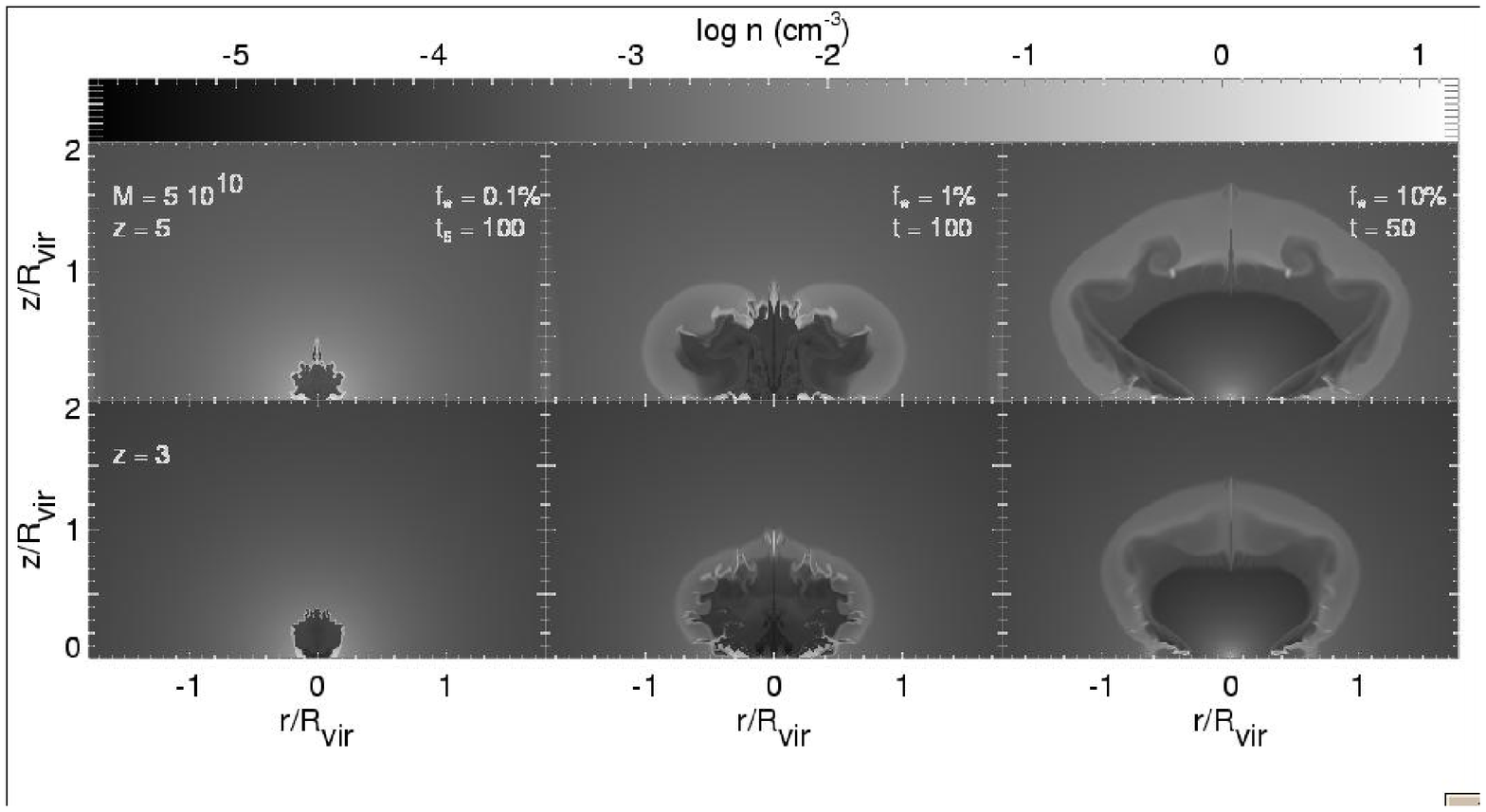}
\caption{
Figure~\ref{figures}{\em c)}
}
\end{figure}

\begin{figure}
\epsscale{1.0}
\hspace*{-1.5cm}
\plotone{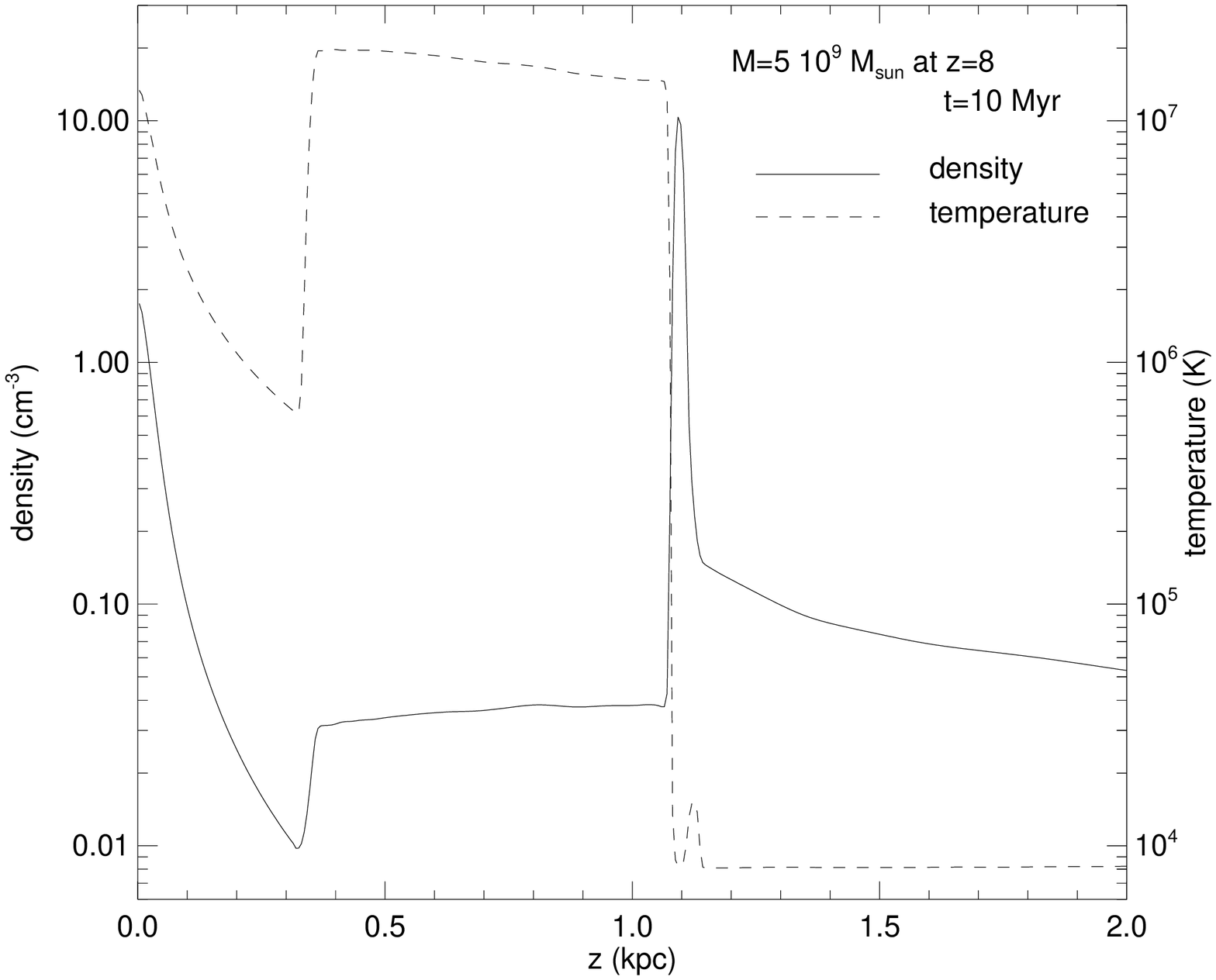}
\caption{Density and temperature distributions of the bubble in the disk 
with $M=5\times10^{9}$~M$_{\odot}$ at redshift 8, 
at $t=10$ Myr after the onset of starburst with $f_*=0.01$. 
We plot them through the vertical z-direction from the galactic center. 
\label{tempden}
}
\end{figure}

\begin{figure}
\epsscale{1.0}
\plotone{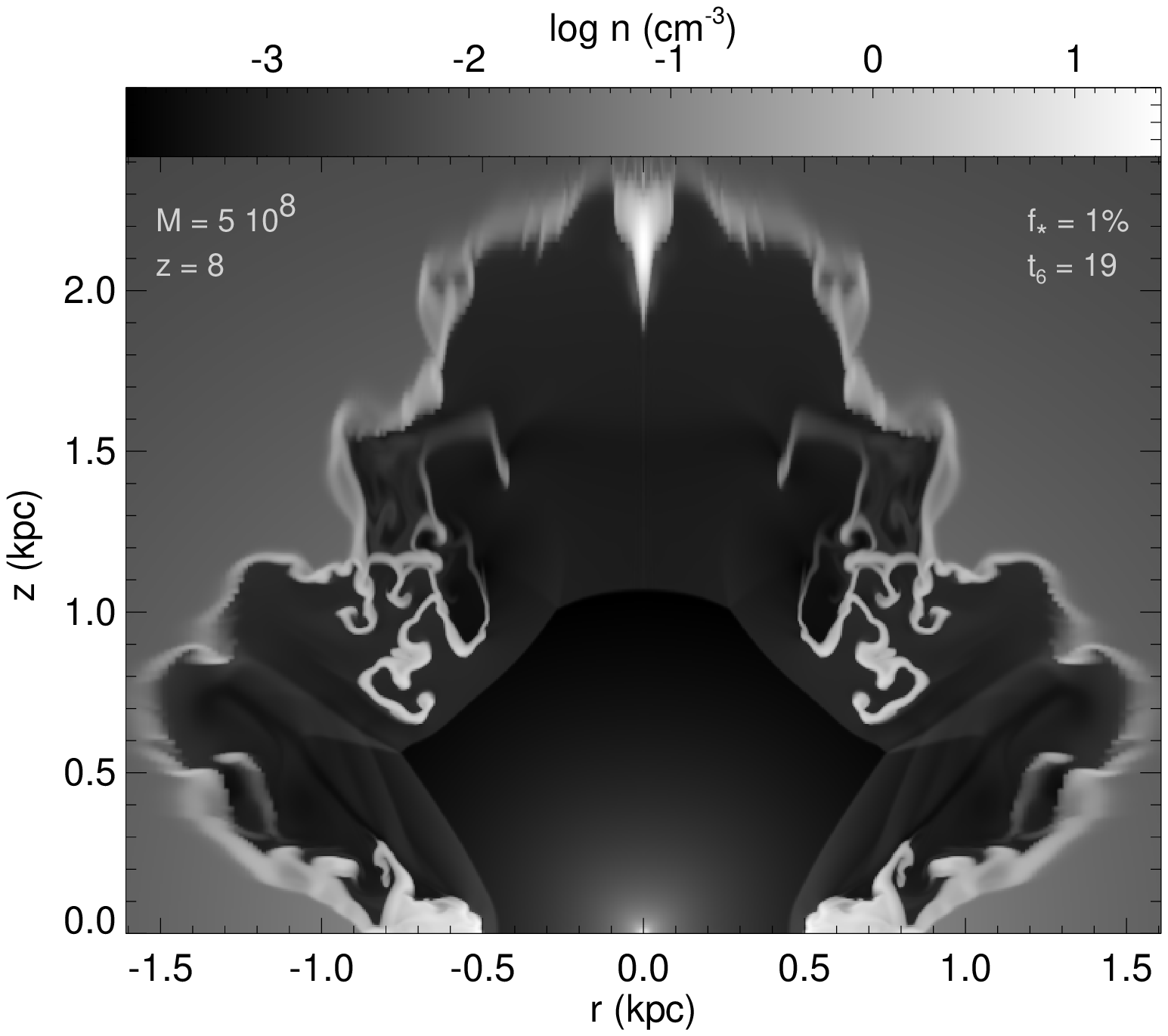} \figcaption{Density
distribution of model with halo mass $M_h=5\times10^{8}$ M$_{\odot}$
at $z=8$ right after the bubble blows out of the disk at $t=19$
Myr with $f_*=1\%$. The simulation was run with a resolution of 2.6 pc within the
central 1 kpc$^{2}$, decreasing smoothly to 26 pc beyond $r\approx
\mbox{z}\approx 2$ kpc. The cooled, swept-up shell of ambient gas fragments
due to Rayleigh-Taylor instability, and mixes with the hot,
low-density interior gas very effectively.
\label{metalmix}
}
\end{figure}

\begin{figure}
\epsscale{1.0}
\plotone{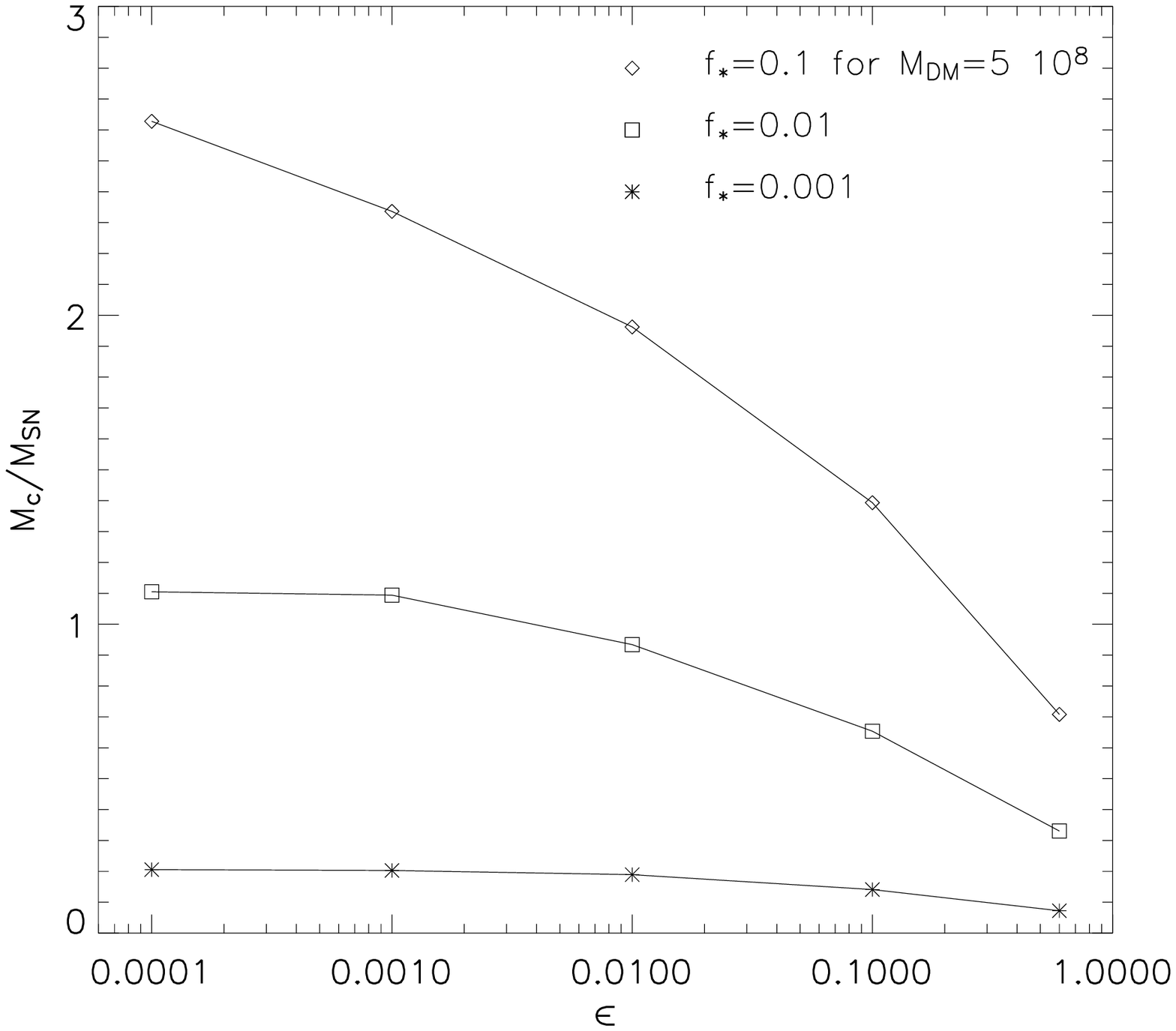}
\figcaption{Figure shows the colored mass $M_c$ divided by the total colored 
mass injected with $c=1$ $M_{SN}$ as a function of 
$\epsilon=1-c$, for the halo with $M_h=5\times10^8$ M$_{\odot}$ at $z=8$ for
star formation efficiencies $f_*=0.001$, 0.01, and 0.1. The masses $M_c$
were measured when the bubbles reached $2R_v$ at $t=100$ Myr when $f_*=0.001$ 
and 0.01 and at $t=50$ Myr when $f_*=0.1$. 
The values of $M_c$ are insensitive to the choices of $\epsilon$ 
as long as they are very small, except for the case with $f_*=0.1$ 
because it was measured at earlier time.
\label{cmass}
}
\end{figure}

\begin{figure}
\epsscale{1.0} 
\plotone{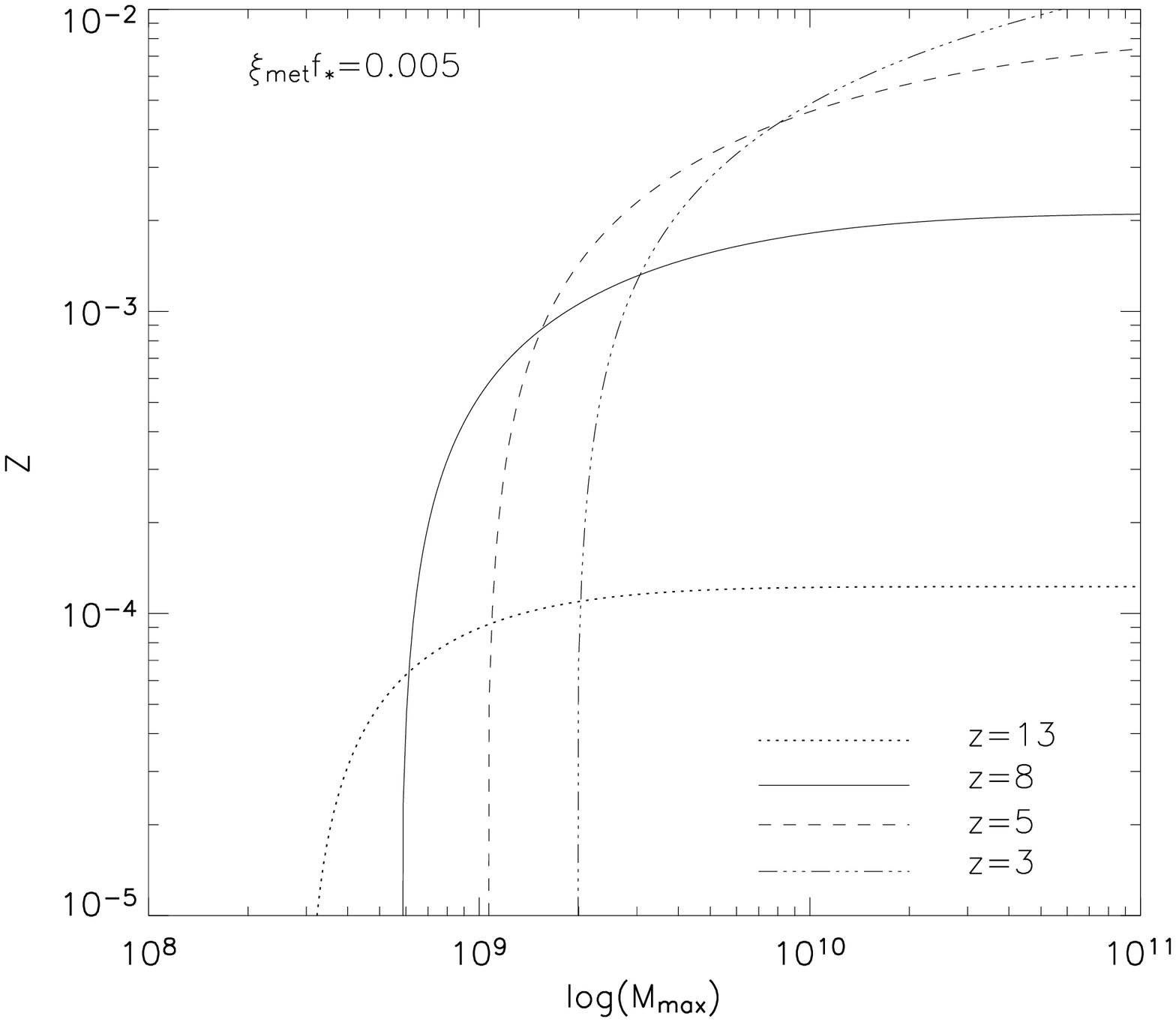} 
\figcaption{
The average metallicity of the universe $\bar{Z}$ from halos with mass between 
$M_{min}(z)$ and $M_{max}$ at redshifts, 3, 5, 8, and 13 for $\xi_{met}f_*=0.005$. 
\label{Zmetal}
}
\end{figure}

\begin{figure}
\epsscale{1.0}
\hspace*{-1.5cm} \plotone{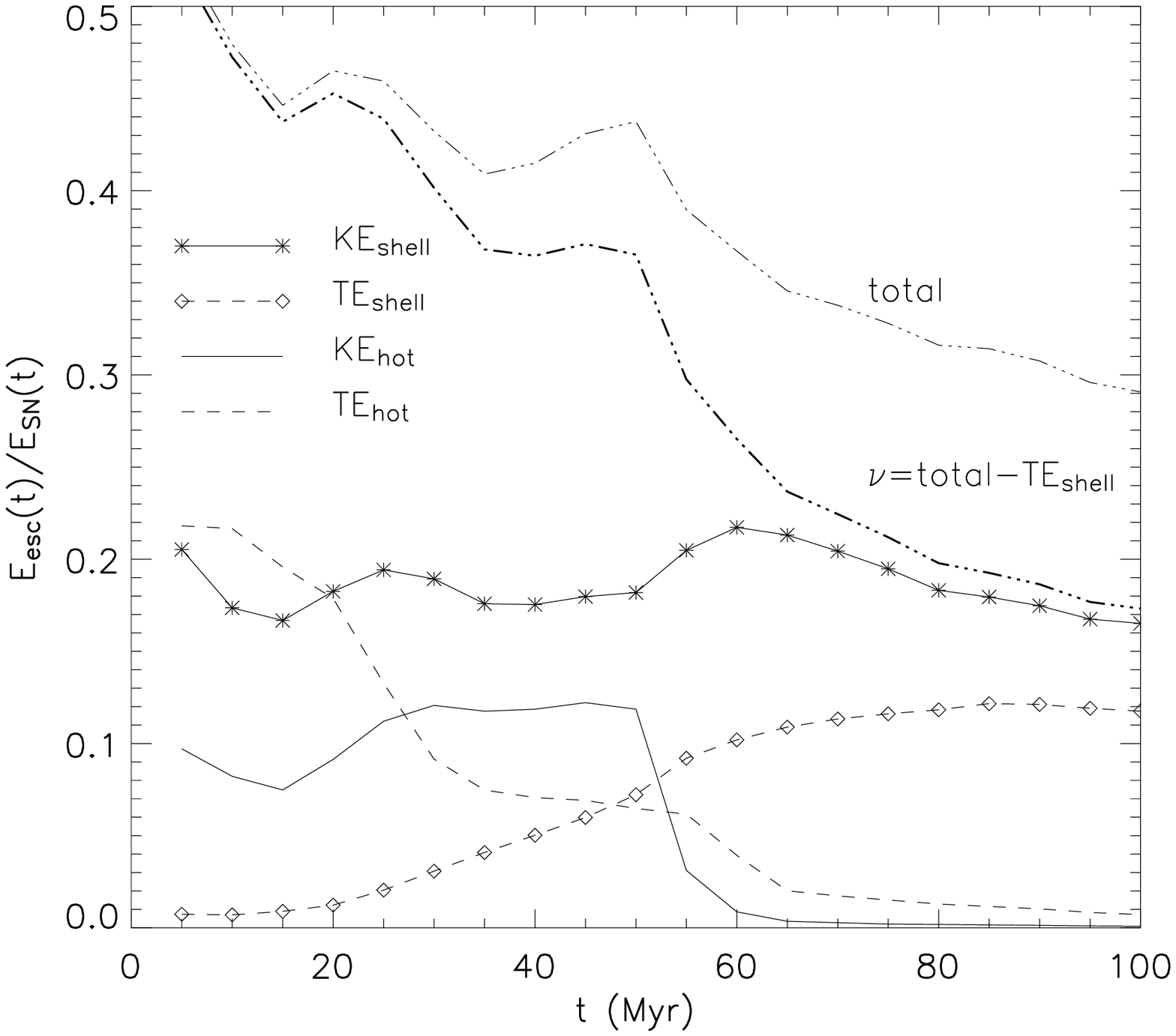} \figcaption{Fractions of total
injected energy escaping from the potential with $M_h=5\times10^{9}$
M$_{\odot}$ at $z=3$: total energy escaping {\em (dash-dot-dot line)},
 total energy excluding thermal energy
carried by swept-up gas liable to be radiated, 
{\cc defined as $\nu$} {\em (thick dash-dot-dot line)}; 
 kinetic energy {\em (solid line with stars)} and thermal
energy {\em (dashed line with diamonds)} carried by swept-up halo gas; and 
 kinetic energy {\em (solid line)} and thermal energy {\em (dashed line)} carried by hot interior
gas with tracer $c<0.6$.
\label{energyesc}
}
\end{figure}

\begin{figure}
\epsscale{1.0}
\hspace*{-1.5cm} \plotone{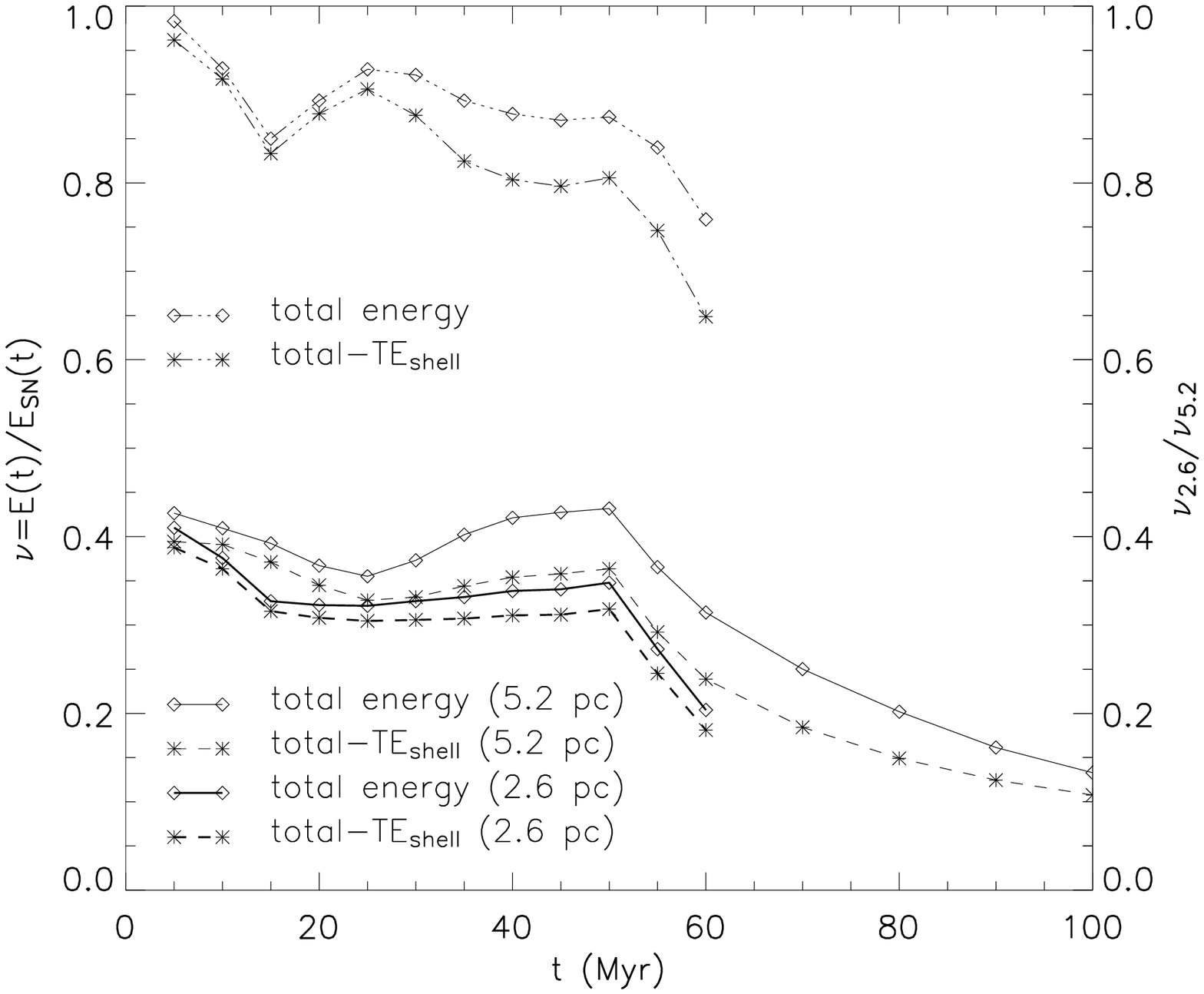} \figcaption{Resolution study
of fractions of total injected energy remaining on the grid with
$M=5\times10^{8}$ M$_{\odot}$ at $z=8$, with $f_*=0.01$. 
The 
main panel shows total energy
{\em (solid line with diamonds)} and total energy excluding the
thermal energy carried by swept-up gas {\em (dashed line with stars)}
with standard resolution (5.2pc: {\em thin lines}) and doubled resolution
(2.6pc: {\em thick lines}).  
The ratio of the results at the two resolutions is shown in the top panel. 
Note that the bubble with doubled resolution
escapes the grid at $t=60$ Myr.
\label{resolution}
}
\end{figure}

\begin{figure}
\epsscale{1.0} 
\plotone{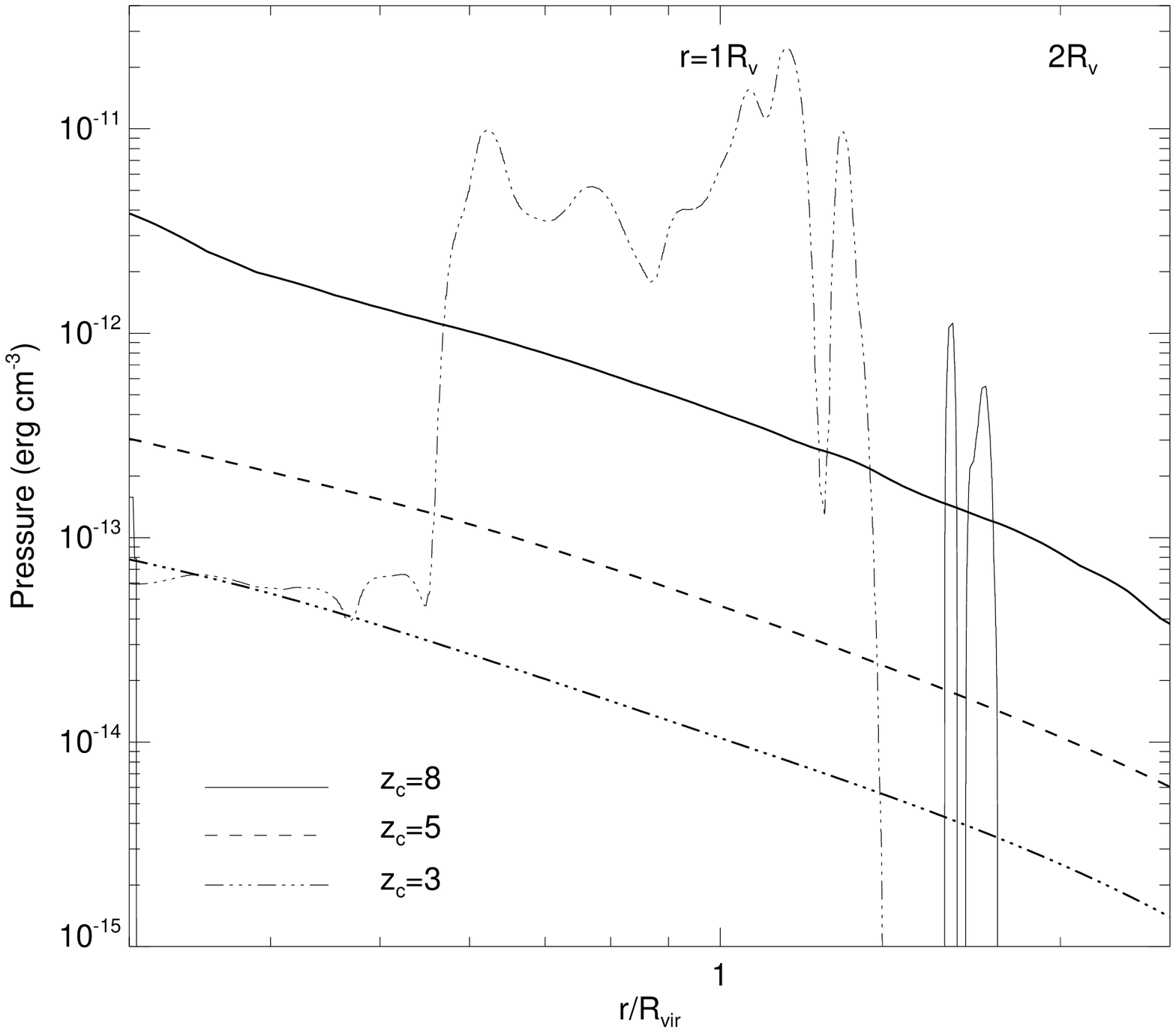} 
\figcaption{
Initial pressure profiles (sum of thermal and ram pressure) for halos
with $M_h=5\times10^{9}$~M$_{\odot}$ at $z=8$ {\em (thick solid
line)}, 5 {\em (thick dashed line)}, and 3 {\em (thick dash-dot-dot line)} as a
function of radius $r$ normalized by virial radius $R_v$. We also plot pressure
distributions of bubbles at 45 degrees from the plane of the disks in the halos
at $z=8$ {\em (solid line)} and 3 {\em (dash-dot-dot line)}, at $t=85$ Myr
after the onset of starbursts with $f_*=0.01$. The bubble at $z=8$ has begun to 
be confined by external pressure, while the bubble at $z=3$ freely expands to the IGM. 
\label{rpg95}
}
\end{figure}

\begin{figure}
\plottwo{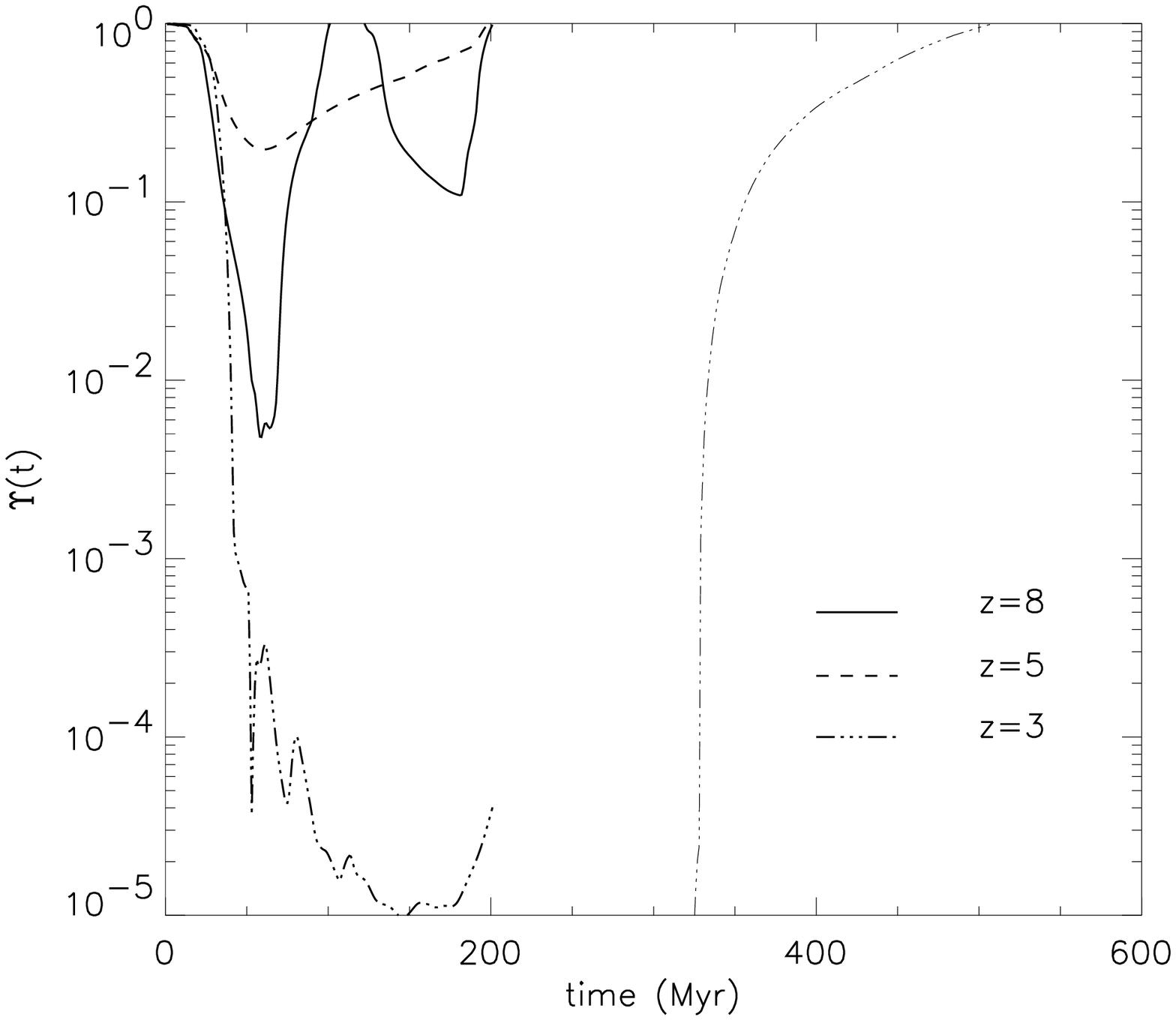}{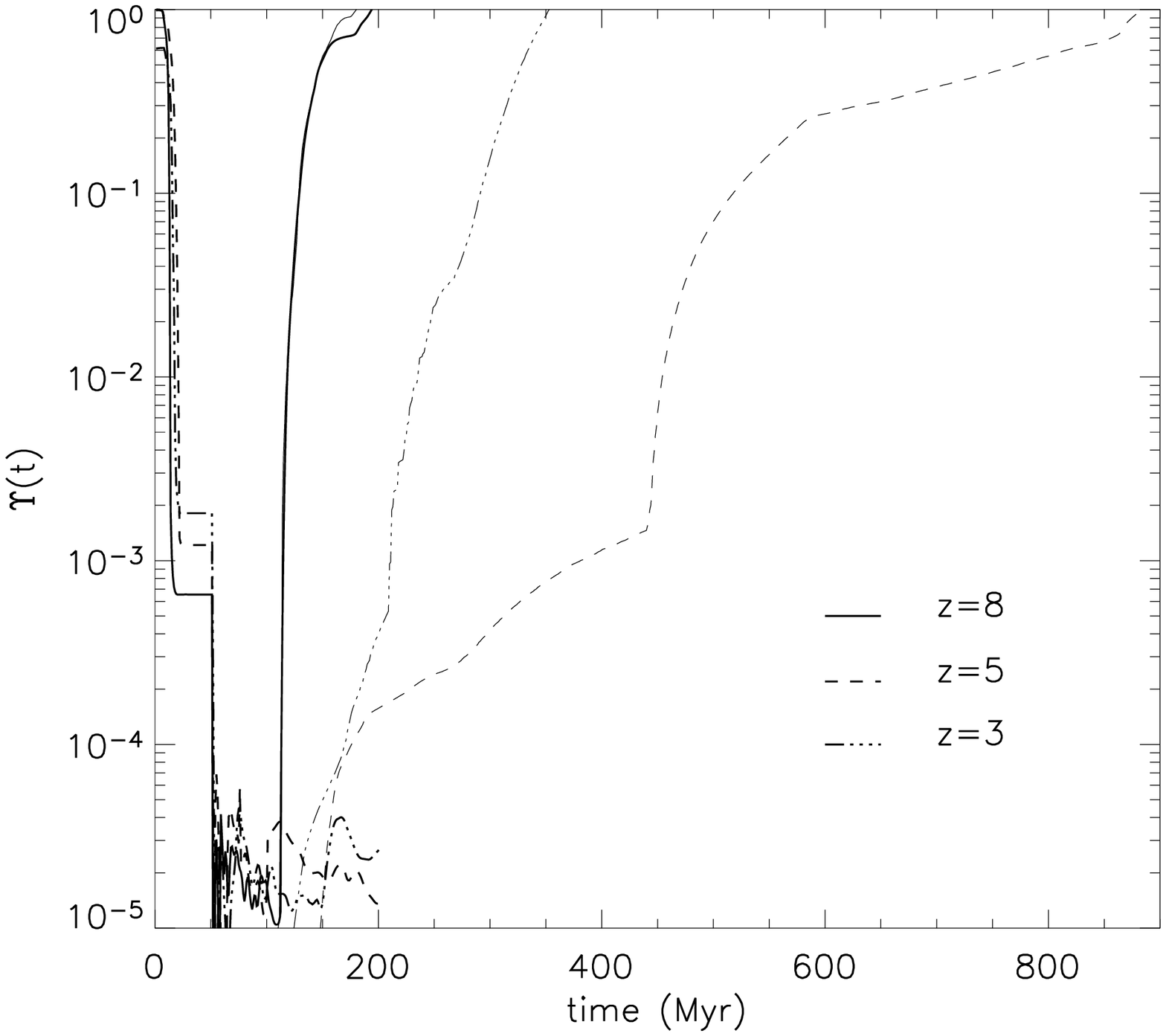} 
\figcaption{
Accretion fraction $\Upsilon(t)$ for halos with $M_h=5\times10^{9}$
M$_{\odot}$ at $z=3$ {\em (dash-dot-dot line)}, 5 {\em (dash line)},
and 8 {\em (solid line)} with star formation efficiency $f_{*}=0.001$
(above) and $0.01$ (below).  The accretion fractions computed from the
simulations are plotted in thick lines and those computed based on the
ballistic approximation are plotted in thin lines.  In the bottom
plot, the ballistic approximation is also compared
with the simulation
directly for the $z=8$ model, showing no more than 5\% deviation.
\label{acm}
}
\end{figure}

\end{document}